\documentclass[longauth]{aa}  

\usepackage{graphicx}
\graphicspath{{./}{figures/}}

\usepackage{caption}
\usepackage{subcaption}

\usepackage[dvipsnames]{xcolor}
\usepackage{txfonts}
\usepackage{makecell}
\usepackage{adjustbox}
\usepackage{threeparttable}
\usepackage{multirow}
\usepackage{dblfloatfix}
\newcommand{\ixpe}{{\it IXPE }}
\newcommand{\ixpeobssim}{{\texttt{ixpeobssim }}}
\newcommand{\I}{{\it I }}
\newcommand{\Q}{{\it Q }}
\newcommand{\U}{{\it U }}
\newcommand{\pd}{{$\Pi$ }}
\newcommand{\pa}{{$\psi$ }}
\newcommand{\pdx}{{$\Pi_{\rm X}$ }}
\newcommand{\pax}{{$\psi_{\rm X}$ }}

\newcommand{\pdo}{{$\Pi_{\rm O}$ }}
\newcommand{\pao}{{$\psi_{\rm O}$ }}

\newcommand{\xmm}{{\it XMM-Newton}}
\newcommand{\swift}{{\it Swift-XRT}}

\def \ergsc{\hbox{erg s$^{-1}$ cm$^{-2}$}}

\def \ergsch{\hbox{erg s$^{-1}$ cm$^{-2}$ Hz$^{-1}$}}

\newcommand{\first}{{\textcolor{magenta}{Obs. 1}}}
\newcommand{\second}{{\textcolor{magenta}{Obs. 2}}}
\newcommand{\third}{{\textcolor{magenta}{Obs. 3}}}
\newcommand{\fourth}{{\textcolor{magenta}{Obs. 4}}}

\def \rotdeg{$\degr/\rm day$}

\usepackage[colorlinks=true, linkcolor=blue, anchorcolor=blue, citecolor=blue, filecolor=blue, menucolor=blue,runcolor=blue, urlcolor=blue, breaklinks=true]{hyperref} 
\usepackage{cleveref}
\usepackage{natbib}
\bibpunct{(}{)}{;}{a}{}{,} 

\begin{document} 
   \title{Magnetic Field Properties inside the Jet of Mrk~421}
   \subtitle{Multiwavelength Polarimetry Including the Imaging X-ray Polarimetry Explorer}
   \author{Dawoon E. Kim \inst{\ref{inst1},\ref{inst2},\ref{inst3}} \thanks{corresponding author, \href{mailto:dawoon.kim@inaf.it}{dawoon.kim@inaf.it}} \orcid{0000-0001-5717-3736}
 \and Laura Di Gesu \inst{\ref{inst4}} \orcid{0000-0002-5614-5028}
 \and Ioannis Liodakis \inst{\ref{inst5}} \orcid{0000-0001-9200-4006}
 \and Alan P. Marscher \inst{\ref{inst6}} \orcid{0000-0001-7396-3332}
 \and Svetlana G. Jorstad \inst{\ref{inst6},\ref{inst7}} \orcid{0000-0001-9522-5453}
 \and Riccardo Middei \inst{\ref{inst8},\ref{inst9}} \orcid{0000-0001-9815-9092}
\and Herman L. Marshall \inst{\ref{inst17}} \orcid{0000-0002-6492-1293}
 \and Luigi Pacciani \inst{\ref{inst1}} \orcid{0000-0001-6897-5996}
 \and Iván Agudo \inst{\ref{inst10}} \orcid{0000-0002-3777-6182}
 \and Fabrizio Tavecchio \inst{\ref{inst11}} \orcid{0000-0003-0256-0995}
 \and Nicolò Cibrario \inst{\ref{inst12},\ref{inst13}} \orcid{0000-0003-3842-4493}
 \and Stefano Tugliani \inst{\ref{inst12}} \orcid{0000-0002-3318-9036}
 \and Raffaella Bonino \inst{\ref{inst13},\ref{inst12}} \orcid{0000-0002-4264-1215}
 \and Michela Negro \inst{\ref{inst14},\ref{inst15},\ref{inst16}} \orcid{0000-0002-6548-5622}
 \and Simonetta Puccetti \inst{\ref{inst8}} \orcid{0000-0002-2734-7835}
 \and Francesco Tombesi \inst{\ref{inst3},\ref{inst18},\ref{inst19}} \orcid{0000-0002-6562-8654}
 \and Enrico Costa \inst{\ref{inst1}} \orcid{0000-0003-4925-8523}
 \and Immacolata Donnarumma \inst{\ref{inst4}} \orcid{0000-0002-4700-4549}
 \and Paolo Soffitta \inst{\ref{inst1}} \orcid{0000-0001-8916-4156}
 \and Tsunefumi Mizuno \inst{\ref{inst20}} \orcid{0000-0001-7263-0296}
 \and Yasushi Fukazawa \inst{\ref{inst21},\ref{inst20},\ref{inst22}} 
 \and Koji S. Kawabata \inst{\ref{inst21},\ref{inst20},\ref{inst22}} 
 \and Tatsuya Nakaoka \inst{\ref{inst20}} 
 \and Makoto Uemura \inst{\ref{inst21},\ref{inst20},\ref{inst22}} 
 \and Ryo Imazawa \inst{\ref{inst21}} 
 \and Mahito Sasada \inst{\ref{inst23}} 
 \and Hiroshi Akitaya \inst{\ref{inst24}} 
 \and Francisco Jos\`{e} Aceituno \inst{\ref{inst10}} 
 \and Giacomo Bonnoli \inst{\ref{inst11},\ref{inst10}} \orcid{0000-0003-2464-9077}
 \and V\`{i}ctor Casanova \inst{\ref{inst10}} 
 \and Ioannis Myserlis \inst{\ref{inst25},\ref{inst26}} \orcid{0000-0003-3025-9497}
 \and Albrecht Sievers \inst{\ref{inst25}} 
 \and Emmanouil Angelakis \inst{\ref{inst27}} 
 \and Alexander Kraus \inst{\ref{inst26}} 
 \and Whee Yeon Cheong \inst{\ref{inst28},\ref{inst29}} \orcid{0009-0002-1871-5824}
 \and Hyeon-Woo Jeong \inst{\ref{inst28},\ref{inst29}} \orcid{0009-0005-7629-8450}
 \and Sincheol Kang \inst{\ref{inst28}} \orcid{0000-0002-0112-4836}
 \and Sang-Hyun Kim \inst{\ref{inst28},\ref{inst29}} \orcid{0000-0001-7556-8504}
 \and Sang-Sung Lee \inst{\ref{inst28},\ref{inst29}} \orcid{0000-0002-6269-594X}
 \and Beatriz Ag\`{i}s-Gonz\`{a}lez \inst{\ref{inst10}} 
 \and Alfredo Sota \inst{\ref{inst10}} \orcid{0000-0002-9404-6952}
 \and Juan Escudero \inst{\ref{inst10}} 
 \and Mark Gurwell \inst{\ref{inst30}} \orcid{0000-0003-0685-3621}
 \and Garrett K. Keating \inst{\ref{inst30}} \orcid{0000-0002-3490-146X}
 \and Ramprasad Rao \inst{\ref{inst30}} 
 \and Pouya M. Kouch \inst{\ref{inst5},\ref{inst31}} \orcid{0000-0002-9328-2750}
 \and Elina Lindfors \inst{\ref{inst5}} 
 \and Ioakeim G. Bourbah \inst{\ref{inst32}} 
 \and Sebastian Kiehlmann \inst{\ref{inst33},\ref{inst32}} 
 \and Evangelos Kontopodis \inst{\ref{inst32}} 
 \and Nikos Mandarakas \inst{\ref{inst33},\ref{inst32}} \orcid{0000-0002-2567-2132}
 \and Stylianos Romanopoulos \inst{\ref{inst33},\ref{inst32}} 
 \and Raphael Skalidis \inst{\ref{inst34},\ref{inst33},\ref{inst32}} 
 \and Anna Vervelaki \inst{\ref{inst32}} \orcid{0000-0003-0271-9724}
 \and Sergey S. Savchenko \inst{\ref{inst7},\ref{inst35},\ref{inst36}} 
 \and Lucio A. Antonelli \inst{\ref{inst9},\ref{inst8}} \orcid{0000-0002-5037-9034}
 \and Matteo Bachetti \inst{\ref{inst37}} \orcid{0000-0002-4576-9337}
 \and Luca Baldini \inst{\ref{inst38},\ref{inst39}} \orcid{0000-0002-9785-7726}
 \and Wayne H. Baumgartner \inst{\ref{inst40}} \orcid{0000-0002-5106-0463}
 \and Ronaldo Bellazzini \inst{\ref{inst38}} \orcid{0000-0002-2469-7063}
 \and Stefano Bianchi \inst{\ref{inst41}} \orcid{0000-0002-4622-4240}
 \and Stephen D. Bongiorno \inst{\ref{inst40}} \orcid{0000-0002-0901-2097}
 \and Alessandro Brez \inst{\ref{inst38}} \orcid{0000-0002-9460-1821}
 \and Niccolò Bucciantini \inst{\ref{inst42},\ref{inst43},\ref{inst44}} \orcid{0000-0002-8848-1392}
 \and Fiamma Capitanio \inst{\ref{inst1}} \orcid{0000-0002-6384-3027}
 \and Simone Castellano \inst{\ref{inst38}} \orcid{0000-0003-1111-4292}
 \and Elisabetta Cavazzuti \inst{\ref{inst4}} \orcid{0000-0001-7150-9638}
 \and Chien-Ting Chen \inst{\ref{inst45}} \orcid{0000-0002-4945-5079 }
 \and Stefano Ciprini \inst{\ref{inst18},\ref{inst8}} \orcid{0000-0002-0712-2479}
 \and Alessandra De Rosa \inst{\ref{inst1}} \orcid{0000-0001-5668-6863}
 \and Ettore Del Monte \inst{\ref{inst1}} \orcid{0000-0002-3013-6334}
 \and Niccolò Di Lalla \inst{\ref{inst46}} \orcid{0000-0002-7574-1298}
 \and Alessandro Di Marco \inst{\ref{inst1}} \orcid{0000-0003-0331-3259}
 \and Victor Doroshenko \inst{\ref{inst47}} \orcid{0000-0001-8162-1105}
 \and Michal Dovčiak \inst{\ref{inst48}} \orcid{0000-0003-0079-1239}
 \and Steven R. Ehlert \inst{\ref{inst40}} \orcid{0000-0003-4420-2838}
 \and Teruaki Enoto \inst{\ref{inst49}} \orcid{0000-0003-1244-3100}
 \and Yuri Evangelista \inst{\ref{inst1}} \orcid{0000-0001-6096-6710}
 \and Sergio Fabiani \inst{\ref{inst1}} \orcid{0000-0003-1533-0283}
 \and Riccardo Ferrazzoli \inst{\ref{inst1}} \orcid{0000-0003-1074-8605}
 \and Javier A. Garcia \inst{\ref{inst50}} \orcid{0000-0003-3828-2448}
 \and Shuichi Gunji \inst{\ref{inst51}} \orcid{0000-0002-5881-2445}
 \and Kiyoshi Hayashida \inst{\ref{inst52}} 
 \and Jeremy Heyl \inst{\ref{inst53}} \orcid{0000-0001-9739-367X}
 \and Wataru Iwakiri \inst{\ref{inst54}} \orcid{0000-0002-0207-9010}
 \and Philip Kaaret \inst{\ref{inst40}} \orcid{0000-0002-3638-0637}
 \and Vladimir Karas \inst{\ref{inst48}} \orcid{0000-0002-5760-0459}
 \and Fabian Kislat \inst{\ref{inst55}} \orcid{0000-0001-7477-0380}
 \and Takao Kitaguchi \inst{\ref{inst49}} 
 \and Jeffery J. Kolodziejczak \inst{\ref{inst40}} \orcid{0000-0002-0110-6136}
 \and Henric Krawczynski \inst{\ref{inst56}} \orcid{0000-0002-1084-6507}
 \and Fabio La Monaca \inst{\ref{inst1}} \orcid{0000-0001-8916-4156}
 \and Luca Latronico \inst{\ref{inst13}} \orcid{0000-0002-0984-1856}
 \and Simone Maldera \inst{\ref{inst13}} \orcid{0000-0002-0698-4421}
 \and Alberto Manfreda \inst{\ref{inst38}} \orcid{0000-0002-0998-4953}
 \and Frédéric Marin \inst{\ref{inst57}} \orcid{0000-0003-4952-0835}
 \and Andrea Marinucci \inst{\ref{inst4}} \orcid{0000-0002-2055-4946}
 \and Francesco Massaro \inst{\ref{inst13},\ref{inst12}} \orcid{0000-0002-1704-9850}
 \and Giorgio Matt \inst{\ref{inst41}} \orcid{0000-0002-2152-0916}
 \and Ikuyuki Mitsuishi \inst{\ref{inst58}} 
 \and Fabio Muleri \inst{\ref{inst1}} \orcid{0000-0003-3331-3794}
 \and C.-Y. Ng \inst{\ref{inst59}} \orcid{0000-0002-5847-2612}
 \and Stephen L. O'Dell \inst{\ref{inst40}} \orcid{0000-0002-1868-8056}
 \and Nicola Omodei \inst{\ref{inst46}} \orcid{0000-0002-5448-7577}
 \and Chiara Oppedisano \inst{\ref{inst13}} \orcid{0000-0001-6194-4601}
 \and Alessandro Papitto \inst{\ref{inst9}} \orcid{0000-0001-6289-7413}
 \and George G. Pavlov \inst{\ref{inst60}} \orcid{0000-0002-7481-5259}
 \and Abel L. Peirson \inst{\ref{inst46}} \orcid{0000-0001-6292-1911}
 \and Matteo Perri \inst{\ref{inst8},\ref{inst9}} \orcid{0000-0003-3613-4409}
 \and Melissa Pesce-Rollins \inst{\ref{inst38}} \orcid{0000-0003-1790-8018}
 \and Pierre-Olivier Petrucci \inst{\ref{inst61}} \orcid{0000-0001-6061-3480}
 \and Maura Pilia \inst{\ref{inst37}} \orcid{0000-0001-7397-8091}
 \and Andrea Possenti \inst{\ref{inst37}} \orcid{0000-0001-5902-3731}
 \and Juri Poutanen \inst{\ref{inst31}} \orcid{0000-0002-0983-0049}
 \and Brian D. Ramsey \inst{\ref{inst40}} \orcid{0000-0003-1548-1524}
 \and John Rankin \inst{\ref{inst1}} \orcid{0000-0002-9774-0560}
 \and Ajay Ratheesh \inst{\ref{inst1}} \orcid{0000-0003-0411-4243}
 \and Oliver Roberts \inst{\ref{inst45}} \orcid{0000-0002-7150-9061}
 \and Roger W. Romani \inst{\ref{inst46}} \orcid{0000-0001-6711-3286}
 \and Carmelo Sgrò \inst{\ref{inst38}} \orcid{0000-0001-5676-6214}
 \and Patrick Slane \inst{\ref{inst30}} \orcid{0000-0002-6986-6756}
 \and Gloria Spandre \inst{\ref{inst38}} \orcid{0000-0003-0802-3453}
 \and Doug Swartz \inst{\ref{inst45}} \orcid{0000-0002-2954-4461}
 \and Toru Tamagawa \inst{\ref{inst49}} \orcid{0000-0002-8801-6263}
 \and Roberto Taverna \inst{\ref{inst62}} \orcid{0000-0002-1768-618X}
 \and Yuzuru Tawara \inst{\ref{inst58}} 
 \and Allyn F. Tennant \inst{\ref{inst40}} \orcid{0000-0002-9443-6774}
 \and Nicholas E. Thomas \inst{\ref{inst40}} \orcid{0000-0003-0411-4606}
 \and Alessio Trois \inst{\ref{inst37}} \orcid{0000-0002-3180-6002}
 \and Sergey S. Tsygankov \inst{\ref{inst31}} \orcid{0000-0002-9679-0793}
 \and Roberto Turolla \inst{\ref{inst62},\ref{inst63}} \orcid{0000-0003-3977-8760}
 \and Jacco Vink \inst{\ref{inst64}} \orcid{0000-0002-4708-4219}
 \and Martin C. Weisskopf \inst{\ref{inst40}} \orcid{0000-0002-5270-4240}
 \and Kinwah Wu \inst{\ref{inst63}} \orcid{0000-0002-7568-8765}
 \and Fei Xie \inst{\ref{inst65},\ref{inst1}} \orcid{0000-0002-0105-5826}
 \and Silvia Zane \inst{\ref{inst63}} \orcid{0000-0001-5326-880X}
 } 
\institute{INAF Istituto di Astrofisica e Planetologia Spaziali, Via del Fosso del Cavaliere 100, 00133 Roma, Italy \label{inst1} 
 \and Dipartimento di Fisica, Universit\`{a} degli Studi di Roma "La Sapienza", Piazzale Aldo Moro 5, 00185 Roma, Italy\label{inst2} 
 \and Dipartimento di Fisica, Universit\`{a} degli Studi di Roma "Tor Vergata", Via della Ricerca Scientifica 1, 00133 Roma, Italy \label{inst3} 
 \and ASI - Agenzia Spaziale Italiana, Via del Politecnico snc, 00133 Roma, Italy\label{inst4} 
 \and Finnish Centre for Astronomy with ESO,  20014 University of Turku, Finland\label{inst5} 
 \and Institute for Astrophysical Research, Boston University, 725 Commonwealth Avenue, Boston, MA 02215, USA\label{inst6} 
 \and Saint Petersburg State University, 7/9 Universitetskaya nab., St. Petersburg, 199034 Russia\label{inst7} 
 \and Space Science Data Center, Agenzia Spaziale Italiana, Via del Politecnico snc, 00133 Roma, Italy\label{inst8} 
 \and INAF Osservatorio Astronomico di Roma, Via Frascati 33, 00078 Monte Porzio Catone (RM), Italy\label{inst9} 
 \and MIT Kavli Institute for Astrophysics and Space Research, Massachusetts Institute of Technology, 77 Massachusetts Avenue, Cambridge, MA 02139, USA\label{inst17} 
 \and Instituto de Astrofísica de Andalucía—CSIC, Glorieta de la Astronomía s/n, 18008 Granada, Spain\label{inst10} 
 \and INAF Osservatorio Astronomico di Brera, Via E. Bianchi 46, 23807 Merate (LC), Italy\label{inst11} 
 \and Dipartimento di Fisica, Università degli Studi di Torino, Via Pietro Giuria 1, 10125 Torino, Italy\label{inst12} 
 \and Istituto Nazionale di Fisica Nucleare, Sezione di Torino, Via Pietro Giuria 1, 10125 Torino, Italy\label{inst13} 
 \and University of Maryland, Baltimore County, Baltimore, MD 21250, USA\label{inst14} 
 \and NASA Goddard Space Flight Center, Greenbelt, MD 20771, USA\label{inst15} 
 \and Louisiana State University, Baton Rouge, LA 70803, USA\label{inst16} 
 \and Istituto Nazionale di Fisica Nucleare, Sezione di Roma "Tor Vergata", Via della Ricerca Scientifica 1, 00133 Roma, Italy\label{inst18} 
 \and Department of Astronomy, University of Maryland, College Park, Maryland 20742, USA\label{inst19} 
 \and Hiroshima Astrophysical Science Center, Hiroshima University, 1-3-1 Kagamiyama, Higashi-Hiroshima, Hiroshima 739-8526, Japan\label{inst20} 
 \and Department of Physics, Graduate School of Advanced Science and Engineering, Hiroshima University Kagamiyama, 1-3-1 Higashi-Hiroshima, Hiroshima 739-8526, Japan\label{inst21} 
 \and Core Research for Energetic Universe (Core-U), Hiroshima University, 1-3-1 Kagamiyama, Higashi-Hiroshima, Hiroshima 739-8526, Japan\label{inst22} 
 \and Department of Physics, Tokyo Institute of Technology, 2-12-1 Ookayama, Meguro-ku, Tokyo 152-8551, Japan\label{inst23} 
 \and Planetary Exploration Research Center, Chiba Institute of Technology 2-17-1 Tsudanuma, Narashino, Chiba 275-0016, Japan\label{inst24} 
 \and Institut de Radioastronomie Millim\'etrique, Avenida Divina Pastora 7, Local 20, E-18012 Granada, Spain\label{inst25} 
 \and Max-Planck-Institut f\"ur Radioastronomie, Auf dem H\"ugel 69, D-53121 Bonn, Germany\label{inst26} 
 \and Section of Astrophysics, Astronomy \& Mechanics, Department of Physics, National and Kapodistrian University of Athens, Panepistimiopolis Zografos 15784, Greece\label{inst27} 
 \and Korea Astronomy and Space Science Institute, Daedeokdae-ro 776, Yuseong-gu, Daejeon 34055, Republic of Korea\label{inst28} 
 \and University of Science and Technology, Gajeong-ro 217, Yuseong-gu, Daejeon 34113, Republic of Korea\label{inst29} 
 \and Center for Astrophysics | Harvard \& Smithsonian, 60 Garden St, Cambridge, MA 02138, USA\label{inst30} 
 \and Department of Physics and Astronomy, 20014 University of Turku, Finland\label{inst31} 
 \and Department of Physics, University of Crete, 70013, Heraklion, Greece\label{inst32} 
\and Institute of Astrophysics, Foundation for Research and Technology-Hellas, GR-71110 Heraklion, Greece\label{inst33} 
 \and Owens Valley Radio Observatory, California Institute of Technology, MC 249-17, Pasadena, CA 91125, USA\label{inst34} 
 \and Special Astrophysical Observatory, Russian Academy of Sciences, 369167, Nizhnii Arkhyz, Russia\label{inst35} 
 \and Pulkovo Observatory, St.Petersburg, 196140, Russia\label{inst36} 
 \and INAF Osservatorio Astronomico di Cagliari, Via della Scienza 5, 09047 Selargius (CA), Italy\label{inst37} 
 \and Istituto Nazionale di Fisica Nucleare, Sezione di Pisa, Largo B. Pontecorvo 3, 56127 Pisa, Italy\label{inst38} 
 \and Dipartimento di Fisica, Università di Pisa, Largo B. Pontecorvo 3, 56127 Pisa, Italy\label{inst39} 
 \and NASA Marshall Space Flight Center, Huntsville, AL 35812, USA\label{inst40} 
 \and Dipartimento di Matematica e Fisica, Universit\`a degli Studi Roma Tre, Via della Vasca Navale 84, 00146 Roma, Italy\label{inst41} 
 \and INAF Osservatorio Astrofisico di Arcetri, Largo Enrico Fermi 5, 50125 Firenze, Italy\label{inst42} 
 \and Dipartimento di Fisica e Astronomia, Università degli Studi di Firenze, Via Sansone 1, 50019 Sesto Fiorentino (FI), Italy\label{inst43} 
 \and Istituto Nazionale di Fisica Nucleare, Sezione di Firenze, Via Sansone 1, 50019 Sesto Fiorentino (FI), Italy\label{inst44} 
 \and Science and Technology Institute, Universities Space Research Association, Huntsville, AL 35805, USA\label{inst45} 
 \and Department of Physics and Kavli Institute for Particle Astrophysics and Cosmology, Stanford University, Stanford, California 94305, USA\label{inst46} 
 \and Institut f\"ur Astronomie und Astrophysik, Universität Tübingen, Sand 1, 72076 T\"ubingen, Germany\label{inst47} 
 \and Astronomical Institute of the Czech Academy of Sciences, Boční II 1401/1, 14100 Praha 4, Czech Republic\label{inst48} 
 \and RIKEN Cluster for Pioneering Research, 2-1 Hirosawa, Wako, Saitama 351-0198, Japan\label{inst49} 
 \and California Institute of Technology, Pasadena, CA 91125, USA\label{inst50} 
 \and Yamagata University,1-4-12 Kojirakawa-machi, Yamagata-shi 990-8560, Japan\label{inst51} 
 \and Osaka University, 1-1 Yamadaoka, Suita, Osaka 565-0871, Japan\label{inst52} 
 \and University of British Columbia, Vancouver, BC V6T 1Z4, Canada\label{inst53} 
 \and Department of Physics, Faculty of Science and Engineering, Chuo University, 1-13-27 Kasuga, Bunkyo-ku, Tokyo 112-8551, Japan\label{inst54} 
 \and Department of Physics and Astronomy and Space Science Center, University of New Hampshire, Durham, NH 03824, USA\label{inst55} 
 \and Physics Department and McDonnell Center for the Space Sciences, Washington University in St. Louis, St. Louis, MO 63130, USA\label{inst56} 
 \and Université de Strasbourg, CNRS, Observatoire Astronomique de Strasbourg, UMR 7550, 67000 Strasbourg, France\label{inst57} 
 \and Graduate School of Science, Division of Particle and Astrophysical Science, Nagoya University, Furo-cho, Chikusa-ku, Nagoya, Aichi 464-8602, Japan\label{inst58} 
 \and Department of Physics, The University of Hong Kong, Pokfulam, Hong Kong\label{inst59} 
 \and Department of Astronomy and Astrophysics, Pennsylvania State University, University Park, PA 16802, USA\label{inst60} 
 \and Université Grenoble Alpes, CNRS, IPAG, 38000 Grenoble, France\label{inst61} 
 \and Dipartimento di Fisica e Astronomia, Università degli Studi di Padova, Via Marzolo 8, 35131 Padova, Italy\label{inst62} 
 \and Mullard Space Science Laboratory, University College London, Holmbury St Mary, Dorking, Surrey RH5 6NT, UK\label{inst63} 
 \and Anton Pannekoek Institute for Astronomy \& GRAPPA, University of Amsterdam, Science Park 904, 1098 XH Amsterdam, The Netherlands\label{inst64} 
 \and Guangxi Key Laboratory for Relativistic Astrophysics, School of Physical Science and Technology, Guangxi University, Nanning 530004, China\label{inst65} 
 } 

   \date{\today}

  \abstract 
   {}
   {To probe the magnetic field geometry and particle acceleration mechanism in the relativistic jets of supermassive black holes.}
   {We conducted a polarimetry campaign from radio to X-ray wavelengths of the high-synchrotron-peak (HSP) blazar Mrk~421, including Imaging X-ray Polarimetry Explorer (\textit{IXPE}) measurements on 2022 December 6--8. During the \ixpe observation, we also monitored Mrk~421 using \swift{} and obtained a single observation with \xmm\ to improve the X-ray spectral analysis. The time-averaged X-ray polarization was determined consistently based on the event-by-event Stokes parameter analysis, spectropolarimetric fit, and maximum likelihood methods. We examined the polarization variability over both time and energy, the latter via analysis of \ixpe{} data obtained over a time span of 7 months.}
   {We detected X-ray polarization of Mrk~421 with a degree of $\Pi_{\rm X}$=14$\pm$1$\%$ and an electric-vector position angle $\psi_{\rm X}$=107$\pm$3$\degr$ in the 2–8 keV band. From the time variability analysis, we find a significant episodic variation in $\psi_{\rm X}$. During 7 months from the first \ixpe pointing of Mrk~421 in 2022 May, \pax varied across the range of 0$\degr$ to 180$\degr$, while \pdx maintained similar values within $\sim$10--15$\%$. Furthermore, a swing in \pax in 2022 June was accompanied by simultaneous spectral variations. The results of the multiwavelength polarimetry show that the X-ray polarization degree was generally $\sim$2--3 times greater than that at longer wavelengths, while the polarization angle fluctuated. Additionally, based on radio, infrared, and optical polarimetry, we find that rotation of $\psi$ occurred in the opposite direction with respect to the rotation of \pax over longer timescales at similar epochs.}
   {The polarization behavior observed across multiple wavelengths is consistent with previous \ixpe findings for HSP blazars. This result favors the energy-stratified shock model developed to explain variable emission in relativistic jets. We consider two versions of the model, one with linear and the other with radial stratification geometry, to explain the rotation of $\psi_{\rm X}$. The accompanying spectral variation during the \pax rotation can be explained by a fluctuation in the physical conditions, e.g., in the energy distribution of relativistic electrons. The opposite rotation direction of \pa between the X-ray and longer wavelength polarization accentuates the conclusion that the X-ray emitting region is spatially separated from that at longer wavelengths. Moreover, we identify a highly polarized knot of radio emission moving down the parsec-scale jet during the episode of \pax rotation, although it is unclear whether there is any connection between the two events.}
            
\titlerunning{Magnetic Field Properties inside the Jet of Mrk~421}
\authorrunning{Dawoon E. Kim et al.}
   \maketitle
%
\section{Introduction}\label{sec:INTRO}

Relativistic jets from active galactic nuclei (AGN) are the most luminous long-lived phenomena across the entire electromagnetic spectrum in the universe. This characteristic of jets makes them natural laboratories that can be studied through multiwavelength and multimessenger observations from radio to $\gamma$-rays. 

Blazars are a subclass of AGN in which a relativistic plasma jet, propelled from the vicinity of a supermassive black hole, is aligned closely with our line of sight \citep[e.g.][]{2019NewAR..8701541H}. The emission from the jet is relativistically boosted toward us, and therefore dominates the spectral energy distribution (SED). The SED typically displays two broad nonthermal radiation components. The lower-frequency component is generally ascribed to be synchrotron emission from relativistic electrons. On the other hand, interpretation of the higher-frequency component falls into two different scenarios: leptonic and hadronic models. In the case of the leptonic model, the high-energy emission is attributed to Compton scattering of synchrotron photons \citep[synchrotron self-Compton, or SSC; e.g.,][]{1974ApJ...188..353J,1992ApJ...397L...5M} or photons from outside the jet \citep[e.g.,][]{Dermer1993,Sikora1994}. The hadronic scenario explains the high-energy emission as proton synchrotron radiation \citep[e.g.,][]{2000NewA....5..377A,Bottcher2013} or photo-pion production \citep[e.g.,][]{1993A&A...269...67M}.

Blazars are sub-classified into three different types, depending on the frequency of the synchrotron peak of the SED. Low-synchrotron-peaked blazars (LSP) have a peak frequency below $10^{14}~$Hz, intermediate-synchrotron peaked blazars (ISP) between $10^{14}~$Hz and $10^{15}~$Hz, and high-synchrotron peaked blazars (HSPs) above $10^{15}~$Hz \citep{2010ApJ...716...30A}. Thus, among these subclasses, HSPs feature a synchrotron spectrum that smoothly connects from radio to X-ray bands \citep{1998MNRAS.299..433F}. Studies of HSPs therefore provide an opportunity to probe the geometry and dynamics of the magnetic field components present in the synchrotron emission features of the jet.

In this respect, multiwavelength polarimetry can be a prominent diagnostic tool to investigate the particle acceleration processes and the geometrical characteristics of the magnetic field inside the jet \citep[e.g.,][]{1979rpa..book.....R, 2019Galax...7...85Z, 2019Galax...7...20B, 2021Galax...9...37T}. Until recently, polarimetric observations of blazars have been limited to optical and radio bands. However, the successful launch of the Imaging X-ray Polarimetry Explorer in late 2021 \citep[{\textit{IXPE}};][]{weisskopf22} has extended multiwavelength polarimetry up to the X-ray energy band. \ixpe is a joint mission of NASA and the Italian Space Agency (Agenzia Spaziale Italiana, ASI). It carries three identical X-ray telescope systems (detector units; DUs) corresponding to three gas pixel detectors \cite[GPDs,][]{2001Natur.411..662C} and three mirror module assemblies (MMAs). \ixpe measures linear polarization in the 2--8 keV band from the photoelectric track produced by the interaction of X-ray photons and gas inside each GPD. 

Mrk~421 (redshift, $z = 0.030$) is the brightest HSP blazar at X-ray energies, and therefore a prime target for measuring polarization with \ixpe \citep{2019ApJ...880...29L}. Indeed, \ixpe securely detected linear polarization at a level exceeding $10\%$ during a two-day pointing of Mrk~421 in May 2022 \citep{2022ApJ...938L...7D}. In addition, during subsequent observations in June 2022, smooth rotation of the electric-vector position angle \pax was discovered \citep{DiGesu2023}, in contrast to the relatively stable polarization degree and direction found for a few other HSP blazars \citep[e.g.,][]{2022Natur.611..677L, 2023ApJ...953L..28M}. This finding can be interpreted as evidence for a helical magnetic field inside the X-ray emitting portion of the jet. Furthermore, if the X-ray emission, and therefore helical field, occurs mainly in the innermost jet, then it is expected that such rotations of \pax should be routinely observed. If so, X-ray observations can probe the section of the jet where the flow is accelerated and collimated \citep{Marscher2008,Marscher2010}. 

Here we report the results of a new \ixpe observation of Mrk~421, conducted as part of a multiwavelength campaign. In \S\ref{sec:xray}, we present the analysis of the time-averaged data and we search for variability of the polarization properties as a function of time and photon energy. To do this, we compare the finding with the results of the previous three \ixpe observations, and we investigate the polarization variability on long time scales. Next, in \S\ref{sec:Multi}, we report the multiwavelength polarimetry results obtained from simultaneous and quasi-simultaneous measurements. Based on these results, we propose possible geometrical and physical interpretations of the magnetic field geometry inside the jet of Mrk~421 in \S\ref{sec:DISCUSSION}. 

\section{X-ray polarization of Mrk~421}\label{sec:xray}

Previously, \ixpe observed Mrk~421 three times: 2022 May 4-6 (hereafter \first), 2022 June 4--6 (\second), and 2022 June 7--9 (\third) \citep{2022ApJ...938L...7D, DiGesu2023}. Here we focus on an additional observation (Obs. ID: 02004401, \fourth) conducted from UTC 15:30 on 2022 December 6 to 04:00 on December 8, with a $\sim74$ ks net exposure time. During the \ixpe pointing, Mrk~421 was in a normal X-ray activity state \citep[e.g.,][]{2019ApJ...880...29L}, with a 2--8 keV flux of $\sim$2$\times10^{-10}$ \ergsc. From this observation, we have derived the time-averaged polarization degree \pdx and electric-vector position angle \pax over the entire exposure. In order to investigate the X-ray polarization variability, we have also performed time- and energy-resolved analysis.
 
We also observed Mrk~421 with \xmm{} on 2022 December 7 for 7.5 ks. In addition, the Neil Gehrels Swift Observatory (hereby referred to as \swift{}) monitored the blazar to improve the spectral and temporal coverage. Table \ref{tab:xray} presents a summary of the X-ray observations. Detailed information on the data reduction procedures is presented in Appendix \ref{sec:Appendix}. Throughout the manuscript, errors are given at the standard 1$\sigma$ (68\%) confidence level.

\subsection{Time-averaged polarization properties}\label{subsec:xray_pol_time_avg}
 
We have derived the linear polarization parameters from \fourth{} by four different methods: (1) the \citet{2015APh....68...45K} method that is implemented in the \texttt{PCUBE} algorithm in \texttt{ixpeobssim} \citep{2022SoftX..1901194B}, (2) an event-based maximum likelihood method \cite[][ML]{marshall20} that includes simultaneous background estimation, (3) the spectropolarimetric analysis described in \citet{2017ApJ...838...72S} using \texttt{XSPEC}, and (4) a maximum likelihood spectropolarimetric (MLS) fit implemented by the multi nest algorithm. Table \ref{tab:xray} summarizes the \pdx and \pax values obtained from these methods. 

\begin{figure}[t!]
\centering         
\includegraphics[width=0.8\columnwidth]{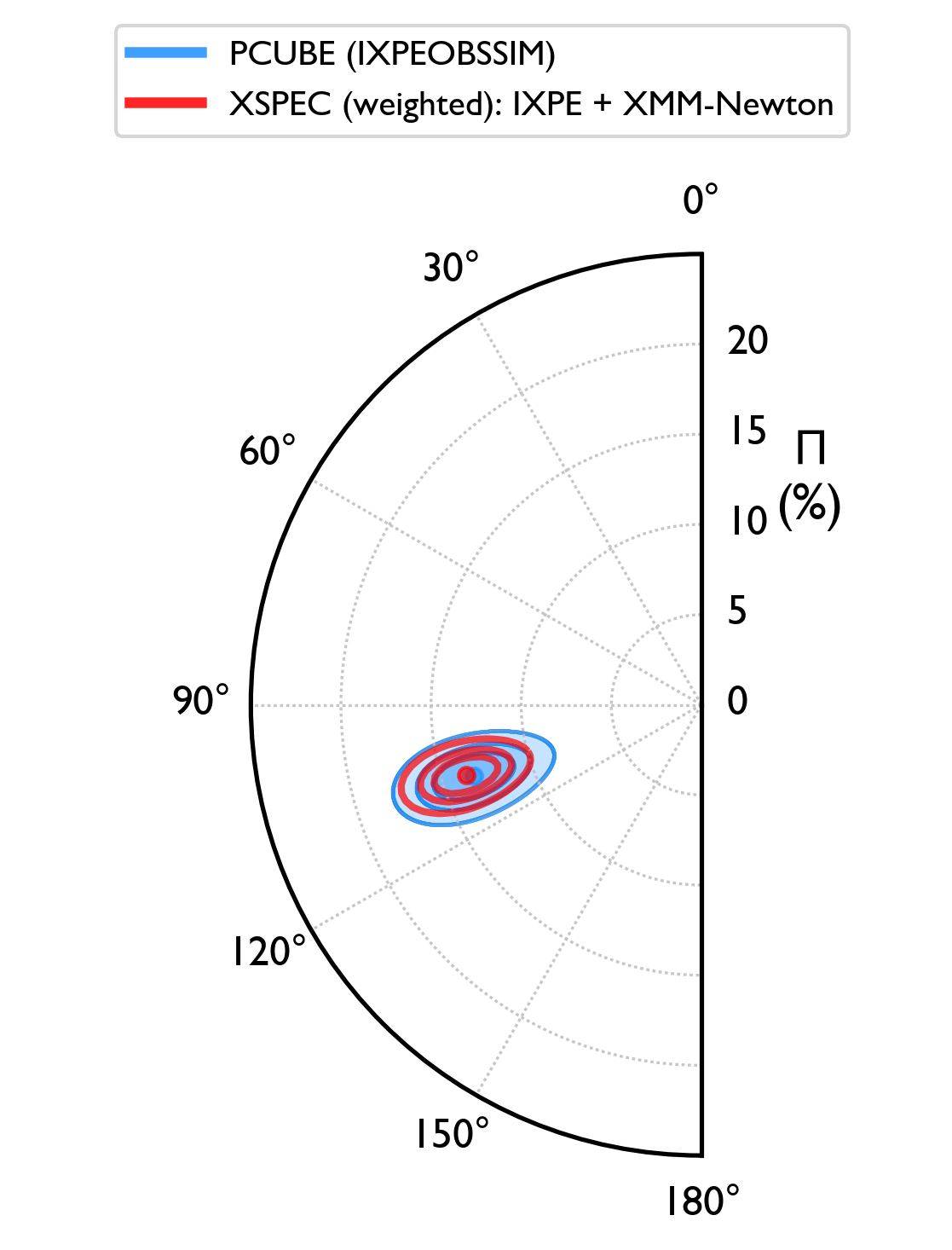}\\
\caption{Polarization contours from \fourth. Contours represent the significance of the time-averaged polarization detected with confidence levels of 68.27$\%$, 90.00$\%$, and 99.00$\%$, with two degrees of freedom. The blue contour indicates the values of \pdx and \pax derived from the \texttt{PCUBE} methods, and the red contours show the same properties from simultaneous \ixpe and {\it XMM-Newton} spectropolarimetric analysis. The radial and angular values represent \pdx and \pax, respectively, with the latter measured from north through east.
}\label{fig:ixpe_polcontour}
\end{figure}

\begin{table*}[ht!]
\caption{Results of the X-ray polarimetric and spectral observations of Mrk~421 \label{tab:xray}}\centering
\begin{tabular}{lllcc}
\hline\hline           
\noalign{\smallskip}
Telescope (Method\footnotemark[1])& Band & Dates & X-ray Polarization & X-ray flux \footnotemark[2] \\
 & (keV) & (YYYY-MM-DD)  & $\Pi_{\rm X}$(\%) \qquad $\psi_{\rm X}$ ($\degr$)  & ($10^{-10}$\ergsc{})\\
\noalign{\smallskip}
\hline
\noalign{\smallskip}
\ixpe{} \fourth{}  (\texttt{PCUBE}) & 2.0 -- 8.0 & 2022-12-06--08 & 13 $\pm$2 \quad  107 $\pm$3 &  --\\
\ixpe{} \fourth{} (\texttt{ML}) & 2.0 -- 8.0 & 2022-12-06--08 & 13 $\pm$1 \quad  107 $\pm$3 & --\\
\ixpe{} \fourth{} (\texttt{XSPEC}) & 2.0 -- 8.0 & 2022-12-06--08 & 14 $\pm$1 \quad  107 $\pm$3 & 2.25 $\pm$0.04\\
\ixpe{} \fourth{} (\texttt{MLS}) & 2.0 -- 8.0 & 2022-12-06--08 & 13 $\pm$1 \quad 109 $\pm$3 & --\\
{\it Swift-XRT} & 0.3 -- 10.0 & 2022-12-06 & -- \qquad\quad -- & 3.20  $\pm$0.06\\
{\it XMM-Newton} & 0.3 -- 10.0 & 2022-12-07 & -- \qquad\quad -- & 2.04 $\pm$0.04 \\
{\it Swift-XRT} & 0.3 -- 10.0 & 2022-12-08 &  -- \qquad\quad -- & 2.47 $\pm$0.05 \\
\noalign{\smallskip}
\noalign{\smallskip}
\hline\noalign{\smallskip}
\ixpe{} \third{} (\texttt{PCUBE})& 2.0 -- 8.0 & 2022-06-07--09 & 10 $\pm$1 \quad Rotation & 3.02 $\pm$0.02\\
\ixpe{} \second{} (\texttt{PCUBE})& 2.0 -- 8.0 & 2022-06-04--06 & 10 $\pm$1 \quad Rotation & 1.57 $\pm$0.01\\
\ixpe{} \first{} (\texttt{PCUBE})& 2.0 -- 8.0 & 2022-05-04--06 & 15 $\pm$2 \quad35 $\pm$4 & 0.87 $\pm$0.003\\
\noalign{\smallskip}
\hline
\end{tabular}
\tablefoot{\footnotemark[1] Methods are described in \S\ref{subsec:xray_pol_time_avg}. \footnotemark[2] X-ray flux in the 2--8 keV band.}
\end{table*}

First, using the \citet{2015APh....68...45K} method, we estimated \pdx=13$\pm$2$\%$ and \pax=107$\pm$3$\degr$ in the 2--8 keV band after background subtraction from the three combined DUs. The \texttt{PCUBE} measurement includes the spectral-model-independent polarization properties over a given energy and time range \citep{2015APh....68...45K}. We derived \pdx and \pax from the normalized Stokes parameters \Q (${\mathcal Q}\equiv\Q/\I$) and \U (${\mathcal U}\equiv\U/\I$), according to {\small $\Pi_{X}$=$\sqrt{({\mathcal Q})^2+({\mathcal U})^2}$} and {\small $\psi_{X}$=$1/2\tan^{-1}({\mathcal U}/{\mathcal Q})$}. Error estimation was calculated based on \citet{2015APh....68...45K} and  \citet{2022arXiv220412739M}. Figure \ref{fig:ixpe_polcontour} indicates the detection significance of this measurement in the form of a polarization contour plot. 

The second method (ML) uses an event-based maximum likelihood method to determine \Q and \U \citep{marshall20}.  The method accounts for background using data from an annulus about the target by including a term in the likelihood, as used in the analysis of the IXPE data of Cen A \citep{Ehlert2022}. Source events were selected from a region of 60\arcsec\ radius about the source centroid, while the background events were selected from an annulus with inner and outer radii of 200\arcsec\ and 300\arcsec. In this case, \pdx= 13$\pm$1$\%$ and \pax=107$\pm$3$\degr$. 

Next, the spectropolarimetric analysis examined the X-ray spectra and polarization properties based on spectral modeling with \texttt{XSPEC} \citep[ver. 12.13.0c;][]{1996ASPC..101...17A}. With this method, we obtained \pdx=14$\pm$1$\%$ and \pax=107$\pm3\degr$. The detection significance of this measurement was $\gtrsim$11$\sigma$. The flux was estimated as 2.25 ($\pm$0.04) $\times10^{-10}\,\ergsc$ over 2--8 keV, the same band over which the \ixpe {\it I}, {\it Q}, and {\it U} spectra were obtained. In this analysis, we additionally included the simultaneous \xmm\ data in order to refine the constraints on the spectral shape over the 0.3--10 keV energy range. The cross-calibration factors among all the spectra are accounted for using the \texttt{CONSTANT} model, normalized to the \xmm\ spectrum while the other spectra were varied. In all the fits, we considered the Galactic absorption along the line of sight of Mrk 421. For this, we used the \texttt{TBABS} model with weighted average column density values from \citet{HI4PI2016} ($N_{H}=1.34\times10^{20} \rm{cm^{-2}}$). The \texttt{WILM} model was applied to take into account metal abundance \citep{2000ApJ...542..914W}. For the spectral modeling, we first applied a simple power law model (\texttt{POWERLAW} in \texttt{XSPEC}) to reproduce the {\it I} synchrotron spectrum, but the best-fit result was poor, with {\tiny$\chi^2/{\mathrm {d.o.f.}}$}=3308/651. Hence, we employed the log-parabolic model (\texttt{LOGPAR}), in which the photon index varies with energy following a log parabola function \citep{2004A&A...413..489M}:
\begin{equation} \label{eq:spectra}
\begin{split}
N(E) = K(E/E_{pivot})^{(\alpha - \beta \log(E/E_{pivot}))},
\end{split}
\end{equation}
\noindent where the pivot energy $E_{pivot}$ is a scaling factor, $\alpha$ describes the slope of the photon spectrum at $E_{pivot}$, $\beta$ expresses the spectral curvature, and $K$ denotes a normalization constant. This spectral model generally describes a typical HSP spectrum well, including that of Mrk~421, both in quiescence and in flaring states \citep{2009ApJ...691L..13D, 2016ApJ...819..156B}. As the photon index varies with energy in this model, the choice of reference energy $E_{pivot}$ changes the determined value of $\alpha$. In our spectral fit, we fixed the pivot energy to 5.0 keV \citep[e.g.,][]{ 2016ApJ...819..156B, 2022MNRAS.514.3179M}. In this case, the $\alpha$ parameter approximately corresponds to the photon index over 3.0--7.0 keV. In our spectral fitting, we also allowed the values of $\alpha$, $\beta$, and $K$ to vary. These free parameters are coupled with the reference spectrum, which here is the {\it XMM-Newton} PN spectrum over the same 2-8 keV band to which the 3 {\it IXPE} DUs are sensitive. Finally, the X-ray polarimetric measurements were constrained by the \texttt{POLCONST} model based on the Stokes parameter fits. Consequently, we obtained statistically acceptable fits, with {\tiny$\chi^2/{\mathrm {d.o.f.}}$}=740/650, above 99$\%$ confidence level. Figure \ref{fig:xspec} and Table \ref{tab:XSPEC} present the parameter values of our best-fit results. We note that in the Stokes-spectra-decoupled case, which only fits {\it I} spectra without involving \texttt{POLCONST}, the best-fit results obtained the same values as we derived from the simultaneous spectropolarimetric fit.

Finally, the MLS method was also applied to determine the polarization properties for the \texttt{POWERLAW} spectral model. With this procedure, we derived the polarization and spectral properties to be \pdx=13$\pm$1$\%$ and \pax=109$\pm$3$\degr$.

We have thus derived that the X-ray polarization properties by four different methods are consistent within the uncertainties (see Table \ref{tab:xray}). The small differences can be explained by the fact that the \texttt{PCUBE} analysis estimated spectral-model-independent polarization properties, while the \texttt{XSPEC} methods take into account the best fit from the spectral modeling. Furthermore, among the current choices, only the \texttt{XSPEC} method can improve sensitivity by applying event weight methods introduced by \citet{2022AJ....163..170D}. Overall, we find that the \pdx and \pax results are robust, with modest statistical errors (see Figure \ref{fig:ixpe_polcontour}).

\subsection{Polarization variability}\label{subsec:pol_variability}
\begin{figure}[t!]
\centering         
\includegraphics[width=1.\columnwidth]{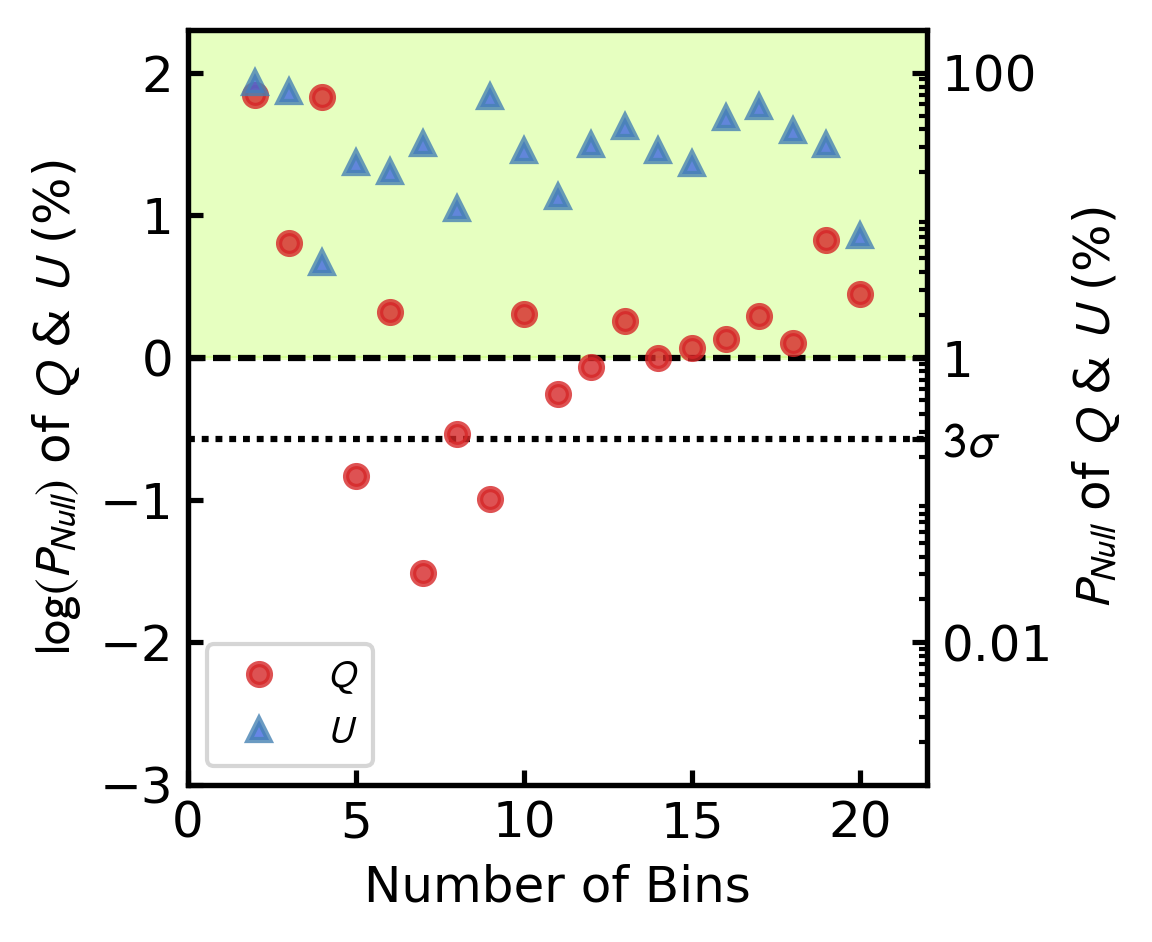}\\
\caption{The null-hypothesis probability of the $\chi^2$ test for time variability of the $\mathcal{Q}$ (red) and $\mathcal{U}$ (blue) Stokes parameters with the constant model for different numbers of time bins for \fourth. The left and right vertical axes correspond to the probability values in logarithmic and linear scales, respectively. The green shaded area indicates that the null-hypothesis probability is above the 1\% significance level. The black dashed and dotted lines located in the middle of the panel represent 1\% and 3$\sigma$ (99.73\%) probability, respectively.
}\label{fig:pvalue}
\end{figure}

\begin{figure}[t!]
\centering         
\includegraphics[width=1.\columnwidth]{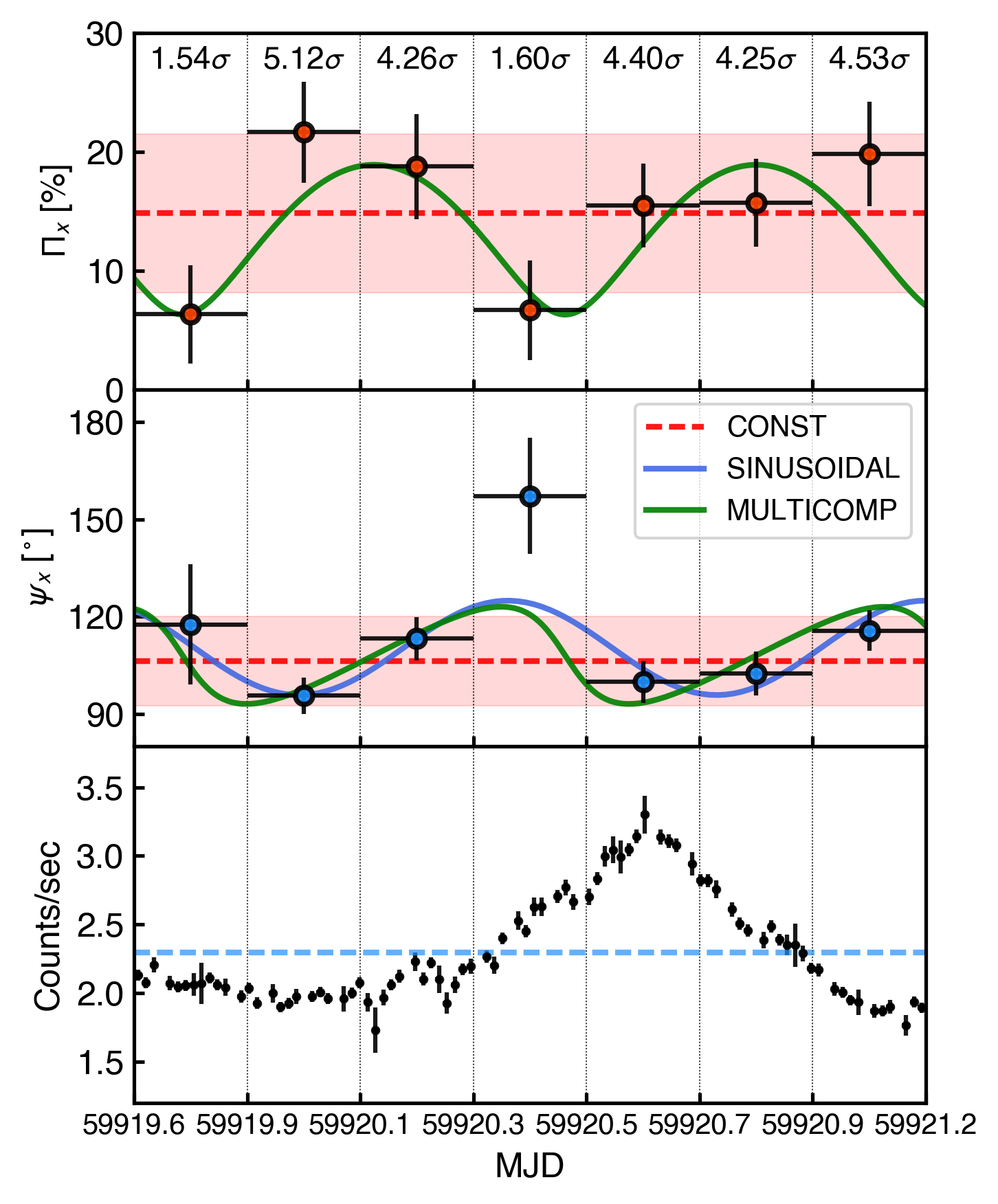}\\
\caption{\ixpe polarization and photon counts versus time during \fourth. From top to bottom: \pdx, \pax, and count rates. Polarization results are from a time-resolved analysis with seven identical 19 ks time bins. The detection significance for each bin is displayed at the top of the figure. In \pdx and \pax panels, red dashed lines denote a fit to a constant function; the shaded area corresponds to $\leq3\sigma$ uncertainty. In addition, the blue and green solid lines indicate the best-fit result with the sinusoidal and multicomponent model, respectively. The blue dashed line in the light curve indicates the average value during the observation.
}\label{fig:LC}
\end{figure}

We have investigated whether the polarization varies as a function of time or energy. We tested the time dependence using two methods. In the first one, we determined the null-hypothesis probability with the $\chi^2$ test using the combined \texttt{PCUBE} and \texttt{XSPEC} analysis \citep[e.g., ][]{2022ApJ...938L...7D, 2023MNRAS.523.4468G, DiGesu2023}. For the second method, we used the unbinned event-based maximum likelihood analysis implemented in \citet{2021AJ....162..134M}. With this method, we can avoid the error that can be caused by subjective selection bias resulting from the binning criteria. 

The first method measured each normalized Stokes parameter value by dividing the data into identical time spans that depended on the selected number of bins (e.g., 2 bins= 100 ks/2 bins = 50 ks/bin). In particular, we split \fourth{} into 2 to 20 time bins. We then compared $\mathcal{Q}$ and $\mathcal{U}$, which followed a Gaussian error distribution, with the results from fitting each parameter treated as constant over time (i.e., $\mathcal{Q}(t)=\mathcal{Q}_{0}$ and $\mathcal{U}(t)=\mathcal{U}_{0}$). The result of fitting the constant model was calibrated for each number of bins. We then calculated the $\chi^2$ and the null-hypothesis probability (for the corresponding degrees of freedom) for each case. Figure \ref{fig:pvalue} indicates the null-hypothesis probability as a function of the number of bins. In this figure, the green-shaded area ($P_{\rm Null}>1\%$) corresponds to cases where the data are statistically consistent with values of  $\mathcal{Q}$ and $\mathcal{U}$ that are constant in time. Conversely, if the points lie outside the region ($P_{\rm Null}<1\%$), it implies that the polarization varies with time. As seen in Figure \ref{fig:pvalue}, we found that splitting the $\mathcal{Q}(t)$ light curves with 5, 7, 8, 9, 11, 12, and 14 time bins does not produce a good fit with the constant model ($P_{\rm Null}<1\%$). The case of 7 time bins has the smallest null-hypothesis probability, with {\tiny$\chi^2/{\mathrm {d.o.f.}}$}=25.26/6, beyond the 3$\sigma$ confidence level. In the polarization light curve of \fourth{} with seven identical time bins (Figure \ref{fig:LC}), we can see that \pax varies from the second to the fourth bin, after which it returns to a more stable value near that of the first three bins. We estimate the rotation rate  of this variation as $\dot{\psi}\sim 92$\rotdeg{} (change by $61\degr$ over $\sim$57 ks from the second to the fourth bin). This result is comparable to the rotation rate of $\dot{\psi}=85$\rotdeg{} reported in \citet{DiGesu2023}. 

Additionally, in order to examine the possibility of continuous change of $\psi_{\rm X}$,  we have tested whether a sinusoidal function ($\psi_{\rm X}(t)$=$A\sin(Ct-D)+B$) provides a better fit of $\psi_{\rm X}(t)$ with respect to a constant function corresponding to the mean value of \pax over 7 time bins (blue line in Figure \ref{fig:LC}). This approach yielded a smaller {\tiny$\chi^2/{\mathrm {d.o.f.}}$} value of 4.75/3, with a Bayesian information criterion (BIC) of 12.53, than the constant model ({\tiny$\chi^2/{\mathrm {d.o.f.}}$}=16.66/6, BIC=18.60). 

\begin{figure}[t!]
\centering         
\includegraphics[width=1.\columnwidth]{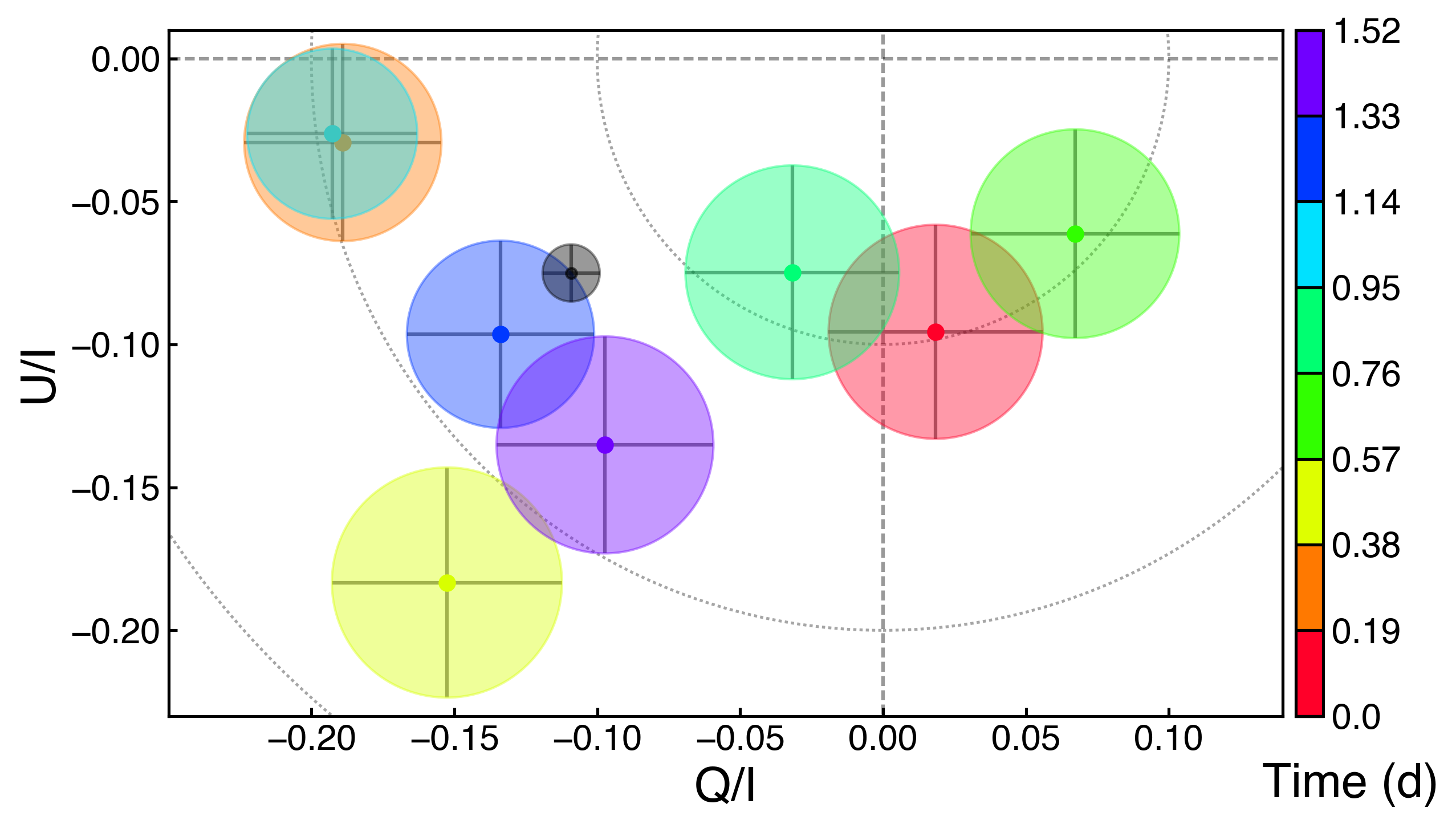}\\
\caption{The time variability of the $\mathcal{Q}$ and $\mathcal{U}$ Stokes parameters during \fourth. Each circle indicates the 1$\sigma$ uncertainties of eight identical time intervals, derived using an event-based maximum likelihood technique as described in \cite{DiGesu2023}. The black circle shows the time-averaged polarization. Each color represents each time interval from the start to the end of the \fourth.
}\label{fig:quplot}
\end{figure}

As a second approach, we employed a maximum likelihood method that allows for \pax rotation in each interval, as described elsewhere \citep{DiGesu2023} and shown in Figure \ref{fig:quplot}. Briefly, the ML method is used (see \S\ref{subsec:xray_pol_time_avg}) for eight equal time intervals but also allowing for \pax rotation during the time interval, as an ``uninteresting'' parameter. This approach prevents \pax rotations during a time interval from reducing the average polarization of that interval, which is important when there are large \pax changes. For example, Figure \ref{fig:quplot} shows that there appear to be large, nearly $90\degr$ \pax shifts between intervals one and two and between intervals three and four. As a result, we confirmed that \pax varied over time during \fourth{} with $\sim 4.5\sigma$ confidence.

In addition, with a maximum likelihood method, we tested a more complex model fit --- the multicomponent model (Pacciani et al. in preparation), which involves the convolution of constant and rotating polarization components --- to check the possibility of continuous variation within the \fourth{} period. The details of this model are described in Appendix \ref{app:multicomp}. From this test, we computed the significance of the multicomponent model relative to the constant polarization model based on the delta likelihood, $\Delta S' = S'(\Pi_0, \Psi_0, 0) - S'(\Pi_{1}, \Psi_{1}, \Pi_{2}, \Psi_{2}, \omega)$. We obtained a decrease in the likelihood estimator, $\Delta L=-32.4$, which follows a chi-square distribution with three degrees of freedom. Hence, the constant polarization model is rejected in favor of the multicomponent model, at the $\sim 5\sigma$ confidence level that this result occurred by chance. The parameters estimated with the multicomponent model are reported in Table \ref{tab:multicomp}. The observed discrepancy between the rotation rate measured from the multicomponent model (i.e., $\omega=245$\rotdeg{}) and the rate estimated directly from the light curve (i.e., $\dot{\psi}\sim$92\rotdeg{}) can be explained by differences in physical interpretation. The multicomponent model assumes a persistent presence of the rotating component in conjunction with the constant emission component, while the other case assumes that the emission originates from a single dominant component. The multicomponent model is indicated with a green line in Figure \ref{fig:LC}.

Therefore, based on these different methods, we conclude that we have found an episodic \pax variation over time during \fourth. Furthermore, the results of sinusoidal and multicomponent models suggest that the variations may be attributed to continuous changes in stable components. However, due to differences in the analysis methods employed, these two models were not compared in this study. Moreover, since these model fit results were derived as relative outcomes compared to the constant model, we cannot rule out the possibility of encountering more complex components and potential stochastic variations.

\begin{table}[t!]
\caption{Parameters for the multicomponent model}\centering
\label{tab:multicomp}
\begin{tabular}{ l l l l l l l }\hline \hline
\noalign{\smallskip}
            $R_1 \cdot\Pi_1$ & $\Psi_1$ & $(1-R_1)\cdot \Pi_2$ & $\Psi_2^0$ & $\omega$\\
            ($\%$)   &  ($\degr$)   &  ($\%$)          & ($\degr$)         & (\rotdeg) \\ \hline \noalign{\smallskip}
            12.6$\pm$1.6 & 108$\pm$ 4& $6.3\pm1.8$& $-96\pm17 $ & $246\pm23$\\      
            \hline
\end{tabular}
\end{table}   

We also tested whether the X-ray polarization depends on energy by applying the same null-hypothesis probability test discussed above, but with the \ixpe energy band (2--8 keV) divided into smaller ranges. We did so for two energy bins (2--4 and 4--8 keV), as well as three energy bins (2--4, 4--6, and 6--8 keV), etc., up to 12 energy bins. We found no statistically significant differences with a constant fit, as $P_{\rm Null}\geq 13\%$ from both $\mathcal{Q}$ and $\mathcal{U}$, hence the data are consistent with energy-independent X-ray polarization. 

\subsection{X-ray polarization and spectral variability across multiple \ixpe observations}\label{subsec:xray_multi}

\begin{figure*}[t!]
     \centering
     \begin{subfigure}{0.24\textwidth}
         \centering
         \includegraphics[width=\textwidth]{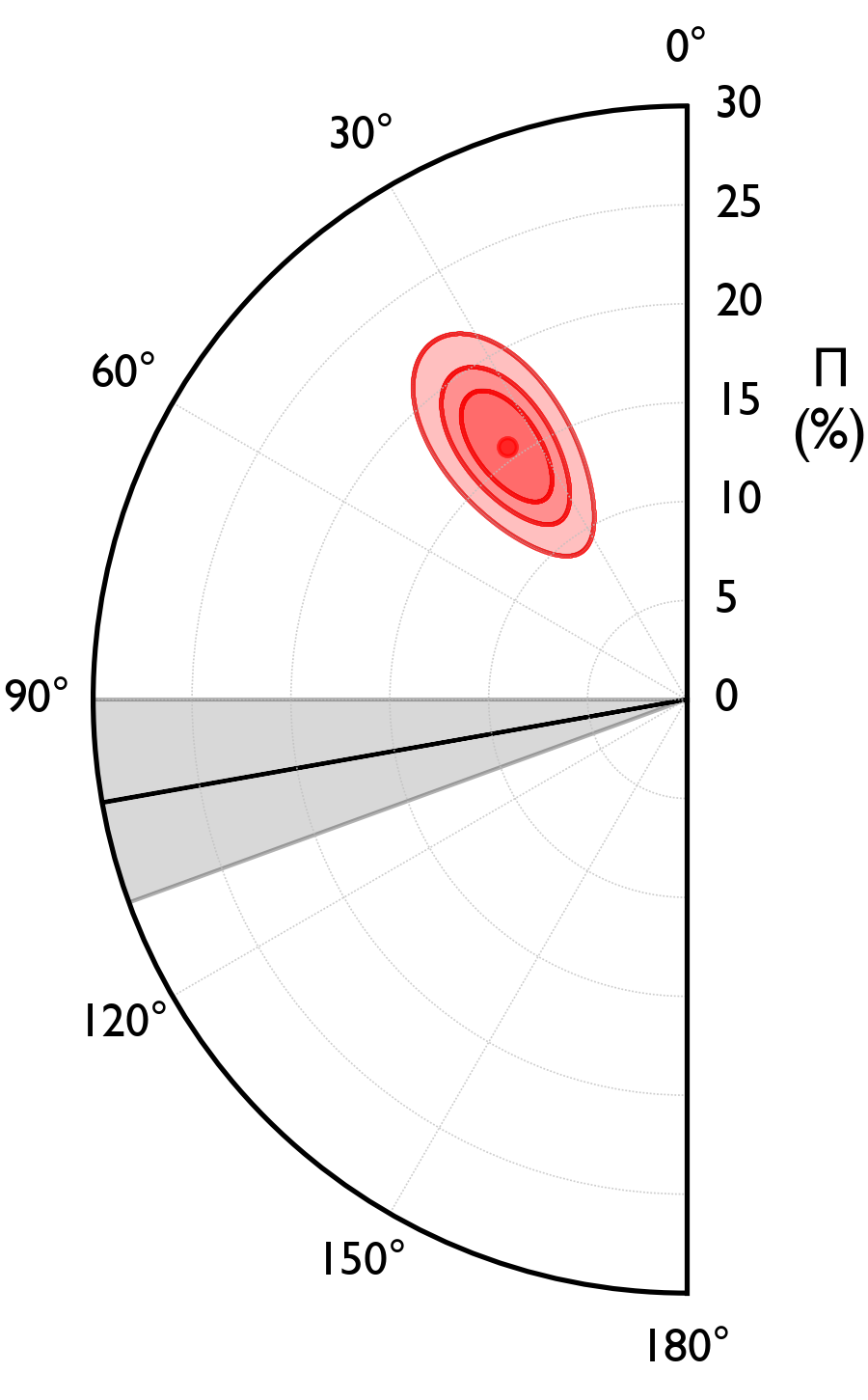}
         \label{First}
     \end{subfigure}
     \hfill
     \begin{subfigure}{0.24\textwidth}
         \centering
         \includegraphics[width=\textwidth]{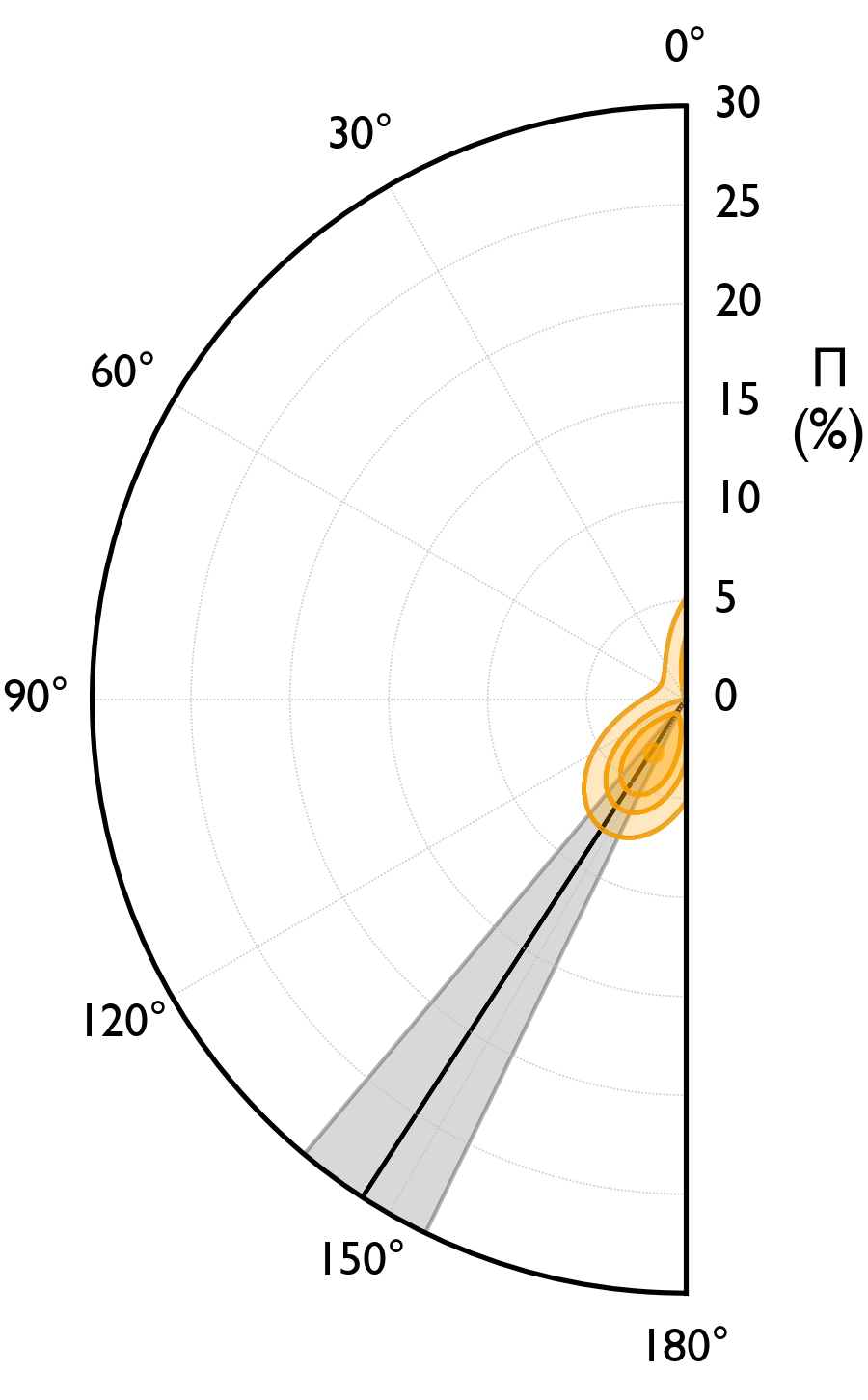}
         \label{Second}
     \end{subfigure}
     \hfill
     \begin{subfigure}{0.24\textwidth}
         \centering
         \includegraphics[width=\textwidth]{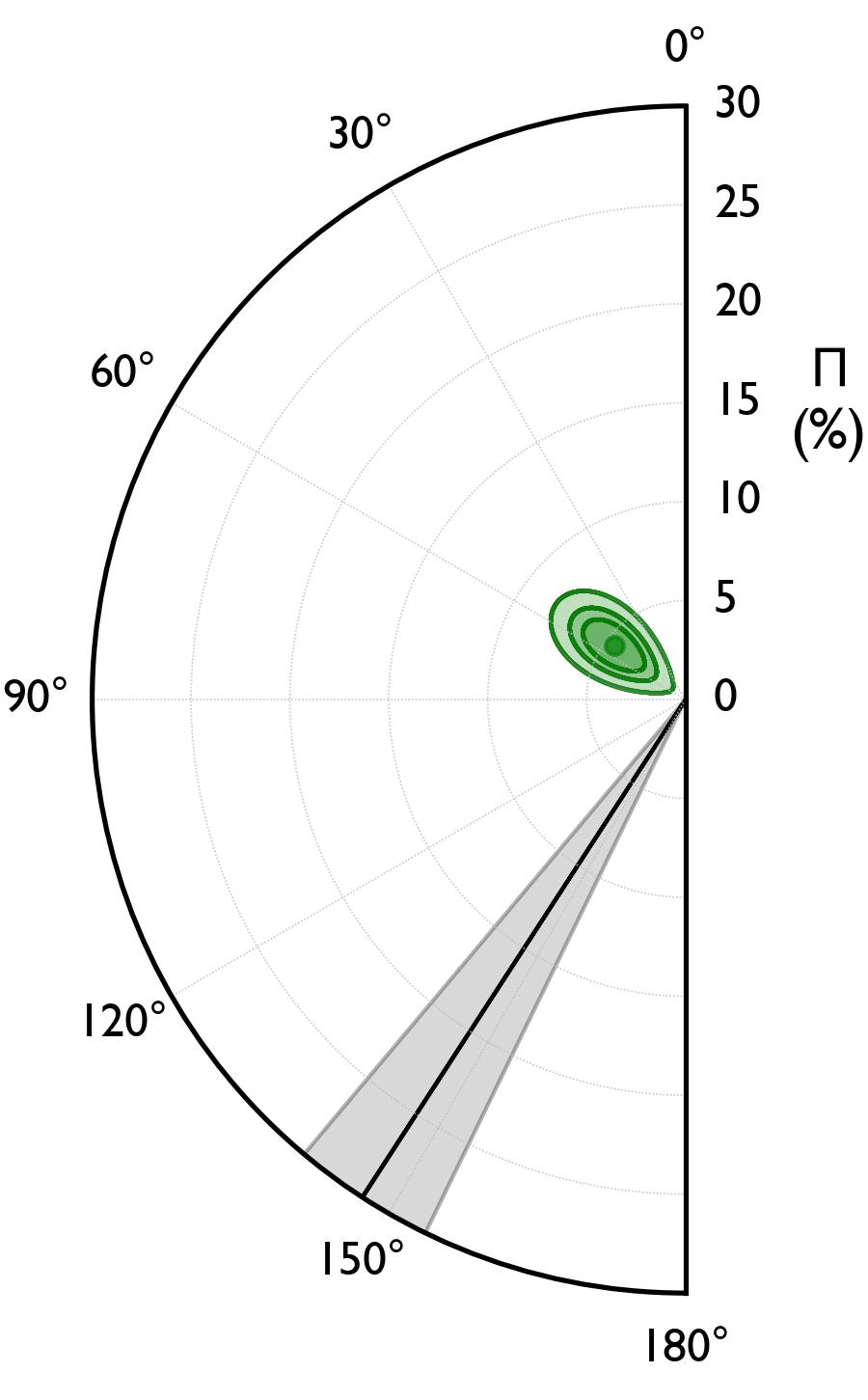}
         \label{Third}
     \end{subfigure}
     \hfill
     \begin{subfigure}{0.24\textwidth}
         \centering
         \includegraphics[width=\textwidth]{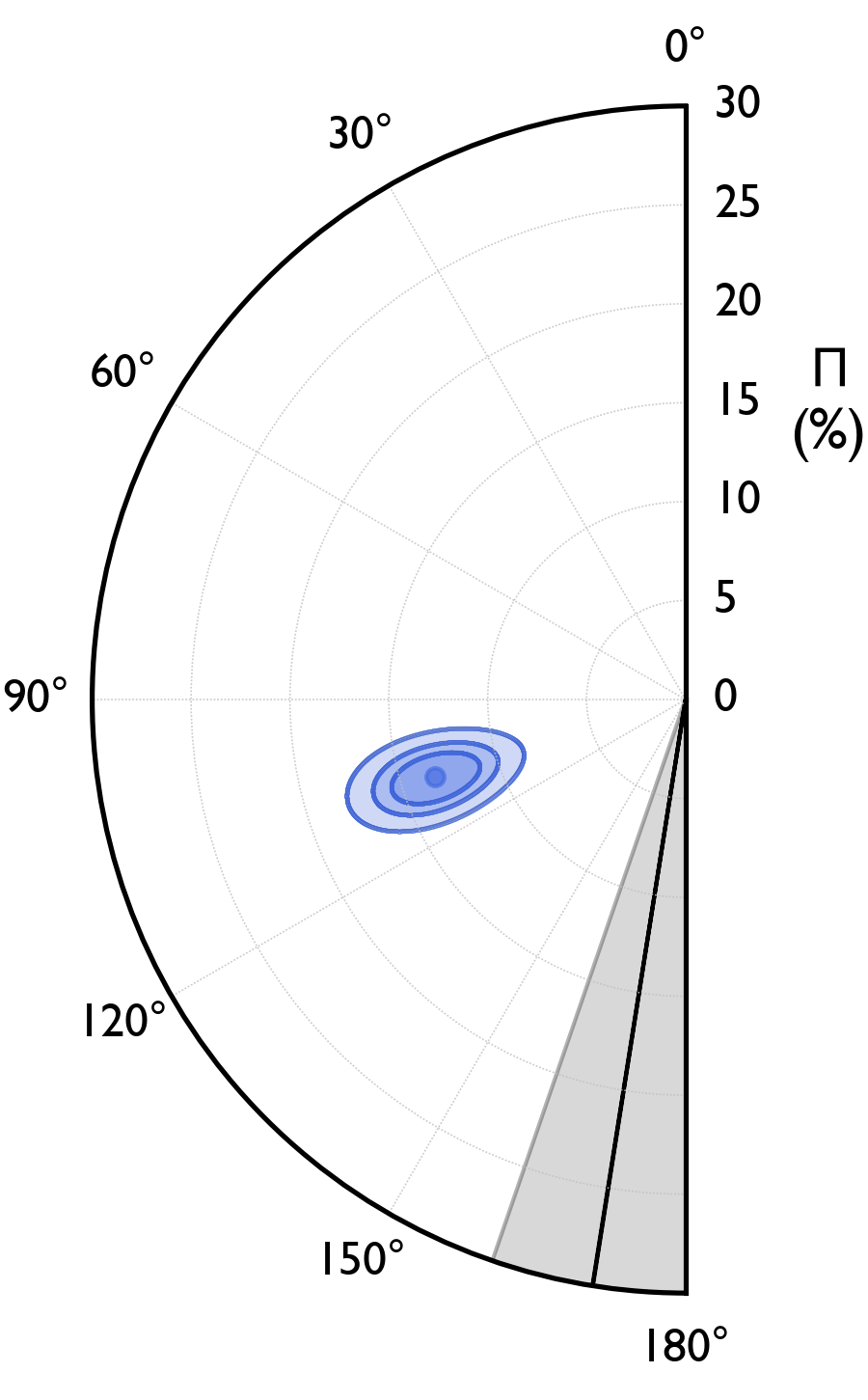}
         \label{Fouth}
     \end{subfigure}
     \vspace{-0.5 cm}
\caption*{Time-averaged polarization contours}
\smallskip\smallskip
          \centering
     \begin{subfigure}{0.24\textwidth}
         \centering
         \includegraphics[width=\textwidth]{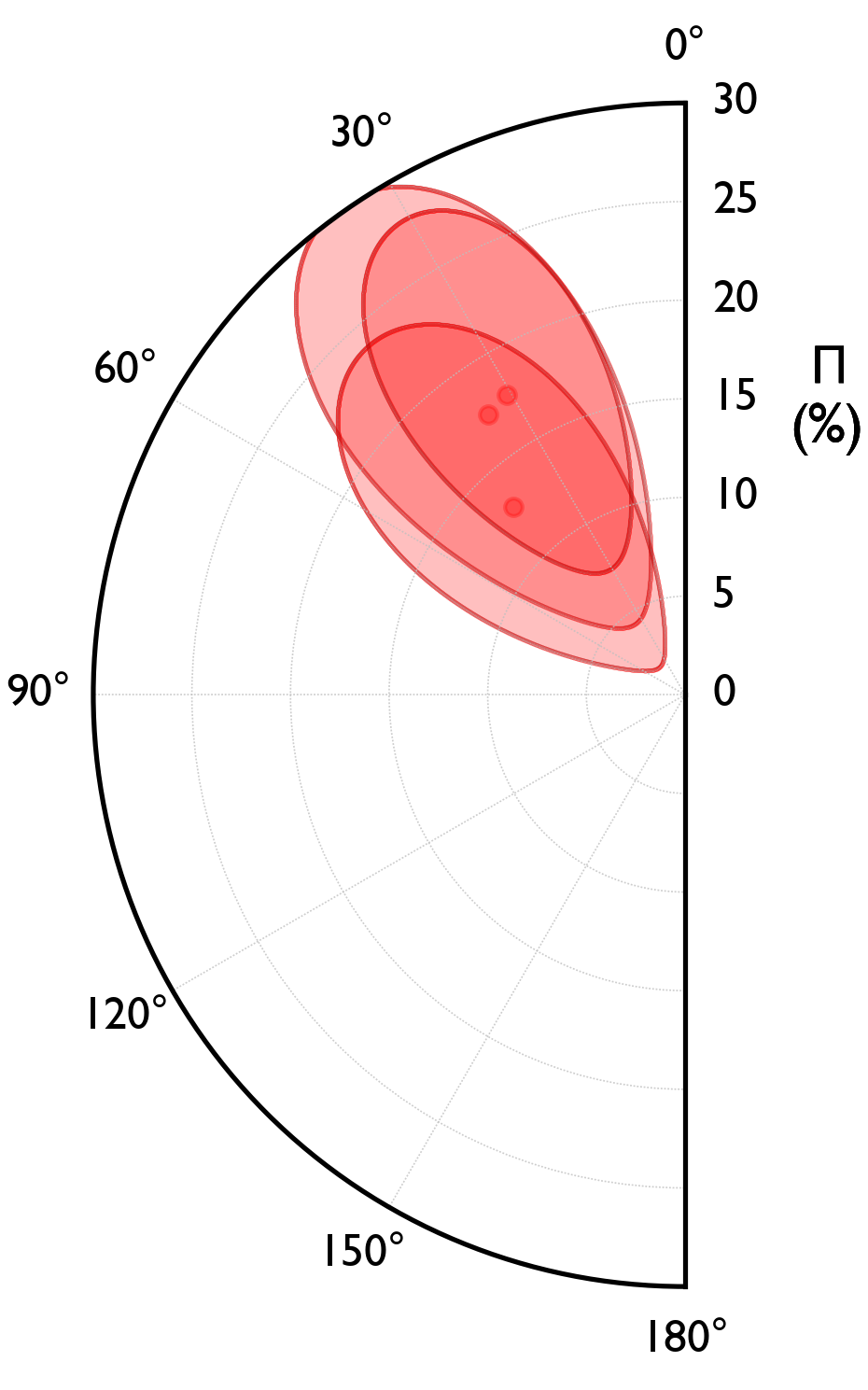}
         \label{first_time}
     \end{subfigure}
     \hfill
     \begin{subfigure}{0.24\textwidth}
         \centering
         \includegraphics[width=\textwidth]{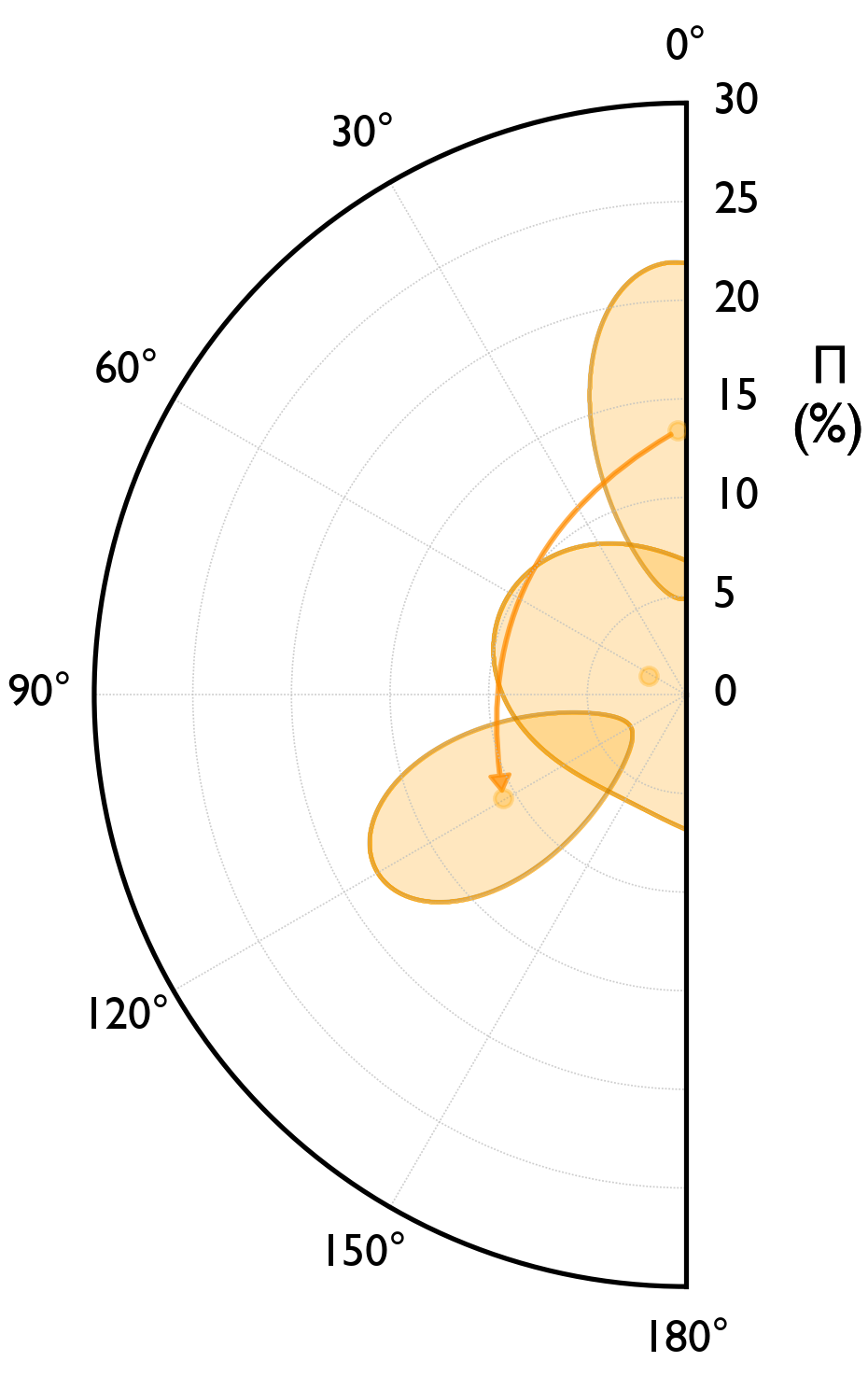}
         \label{second_time}
     \end{subfigure}
     \hfill
     \begin{subfigure}{0.24\textwidth}
         \centering
         \includegraphics[width=\textwidth]{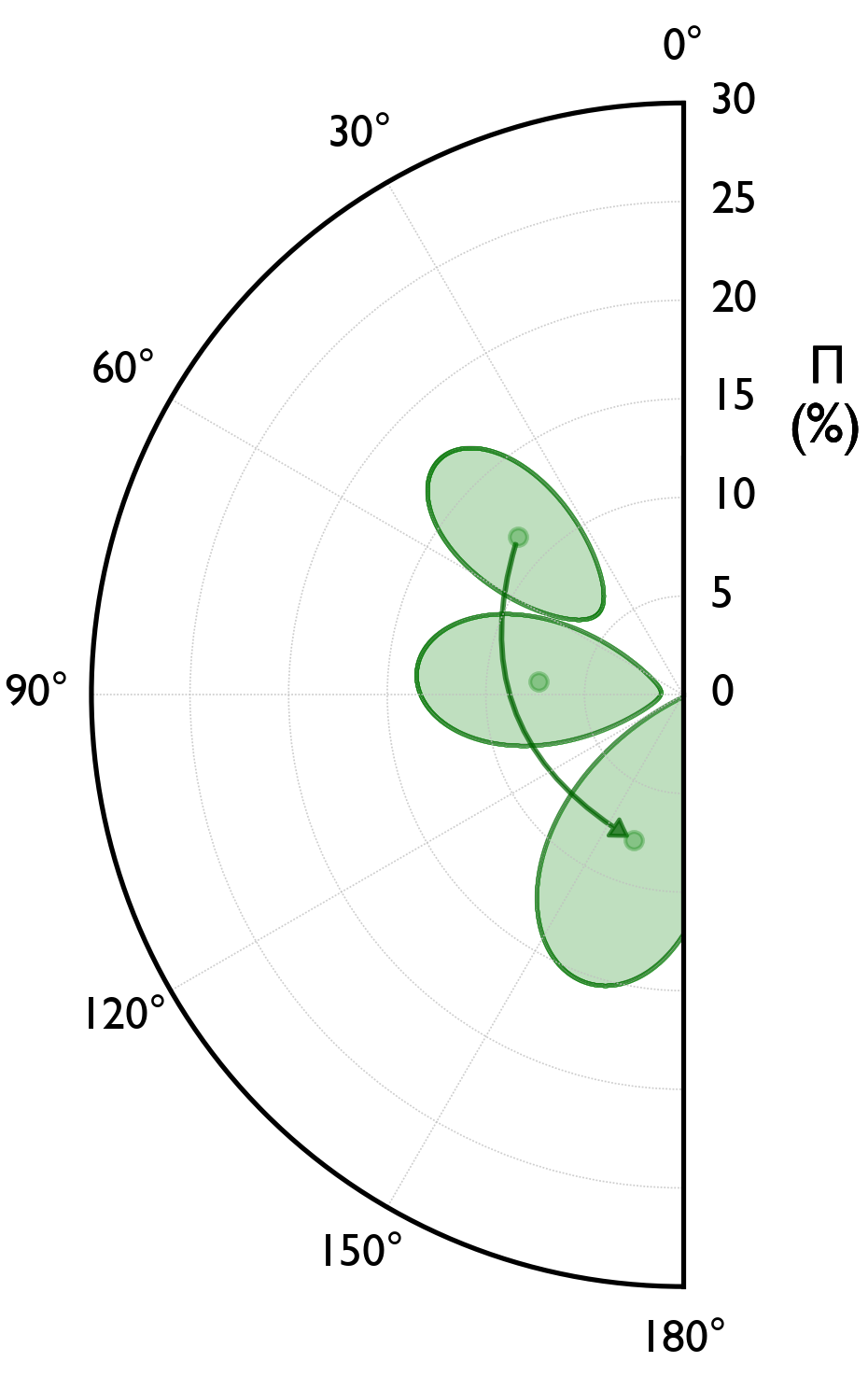}
         \label{third_time}
     \end{subfigure}
     \hfill
     \begin{subfigure}{0.24\textwidth}
         \centering
         \includegraphics[width=\textwidth]{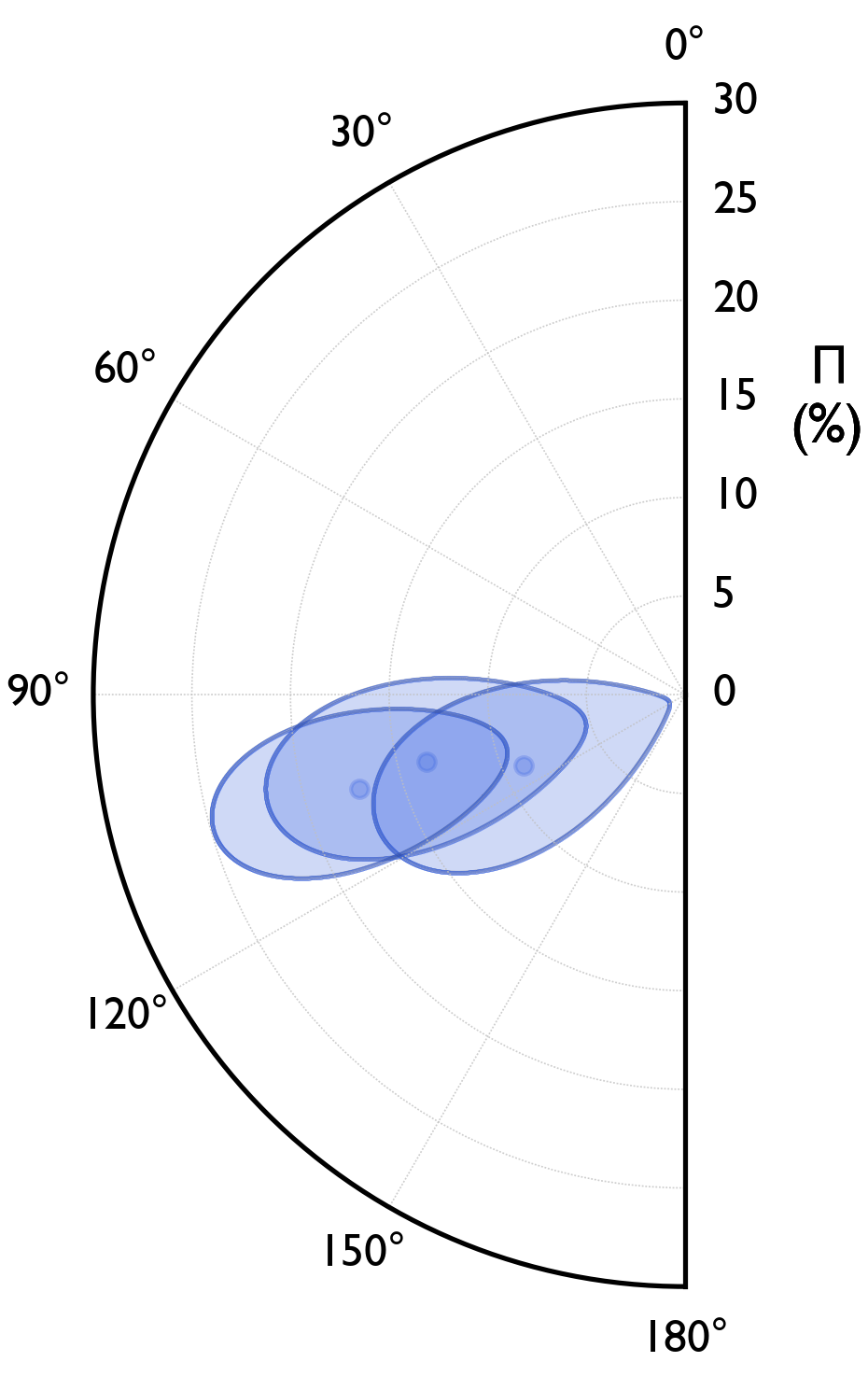}
         \label{fourth_time}
     \end{subfigure}
     \vspace{-0.5 cm}
\caption*{Time-resolved polarization contours}
\smallskip\smallskip
\centering
     \begin{subfigure}{0.5\textwidth}
         \centering
         \includegraphics[width=\textwidth]{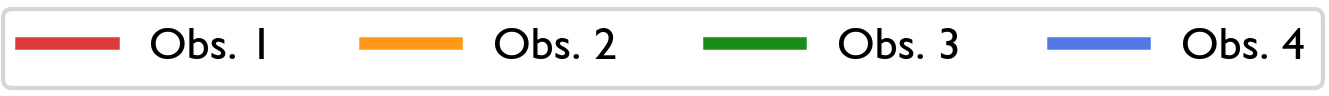}
         \vspace{-0.5 cm}
     \end{subfigure}
\caption{Time-averaged and time-resolved polarization contours of four multiple \ixpe observations. In the top panel, each colored contour represents the significance of the time-averaged polarization detection for the corresponding observation (\first: red, \second: orange, \third: green, and \fourth: blue).  Contours are shown at confidence levels of 68.27\%, 90.00\%, and 99.00\%, from a $\chi^2$ test with two degrees of freedom. Additionally, the black line and gray shaded areas indicate the jet's electric vector position angle (EVPA) in degrees as observed by the VLBA at 43 GHz (\first: 100$\pm$10$\degr$, \second: 147$\pm$7$\degr$, \third: 147$\pm$7$\degr$, and \fourth: 171$\pm$10$\degr$). In the bottom panel, the 99.00\% confidence level contours show the time variation of polarization properties by dividing the data into three identical time intervals for each observation. In the cases of \second\ (orange) and \third\ (green), arrows indicate a rotation of \pax from the start to the end of the observation.
}\label{fig:polcontours}
\end{figure*}

\begin{figure*}[t!]
\centering         
\includegraphics[width=1.\textwidth]{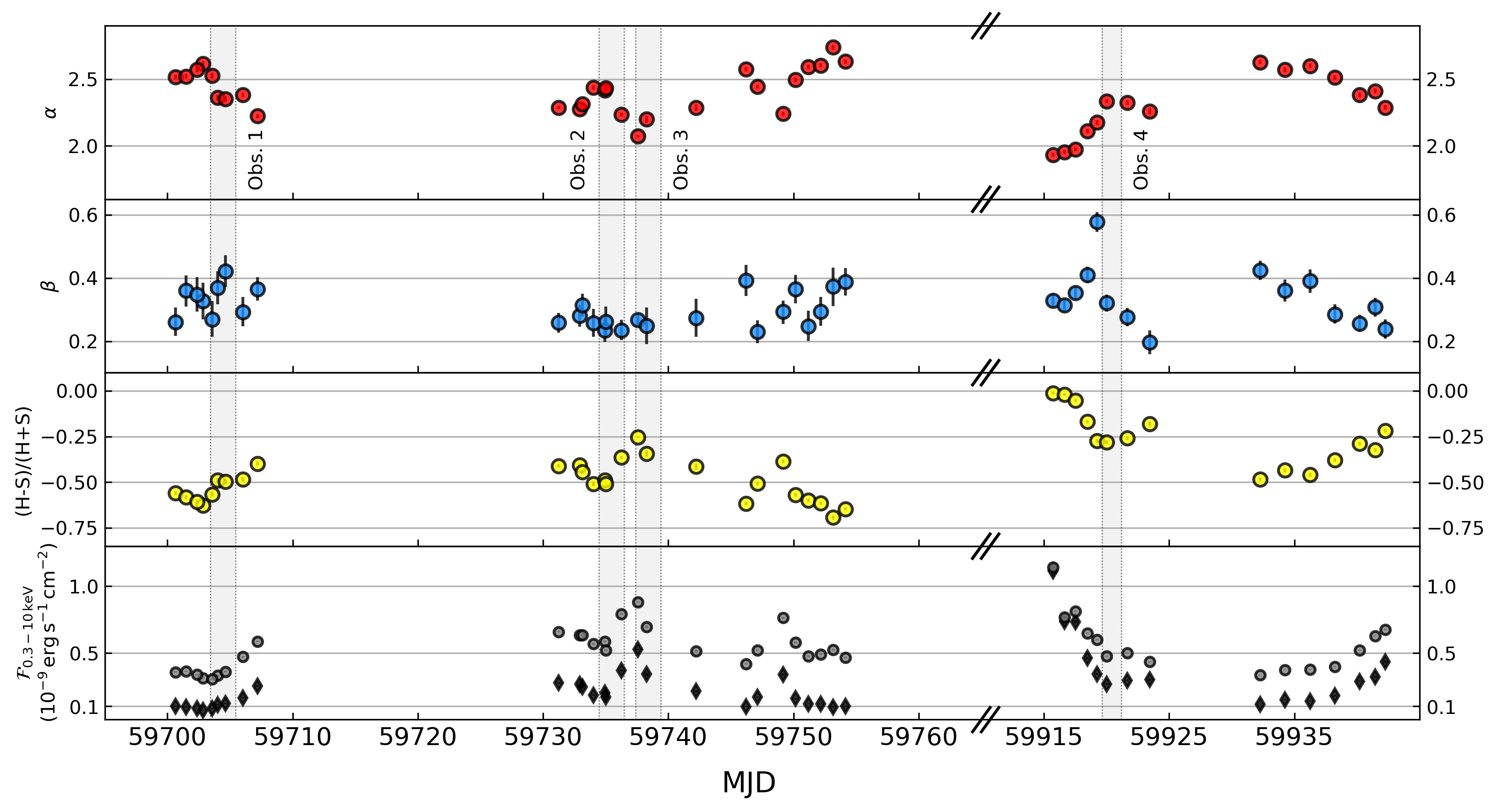}\\
\caption{{\it SWIFT-XRT} X-ray spectral parameters and flux versus time of Mrk~421 from a log-parabolic best-fit model with $E_{pivot}$=5 keV. From top to bottom: $\alpha$, $\beta$, hardness ratio, and flux in the soft (0.3--2 keV, gray circles) and hard (2--10 keV, black diamonds) energy bands. The time ranges of the \ixpe pointings are indicated by the gray-shaded regions.
}\label{fig:swift}
\end{figure*}

Significant X-ray polarization from Mrk~421 has been detected at four epochs with \ixpe (see Table \ref{tab:xray}). Figure \ref{fig:polcontours} shows error contours for both time-averaged and time-resolved data of each observation. Over the seven months from \ixpe \first{} to \fourth, the value of \pax varied widely, with a continuous rotation over 180$\degr$ observed during \second{} and \third. In contrast, measurements of \pdx for all events were similar within a range of 10–15$\%$, even though the time-averaged \pdx of \second{} and \third{} appears to be clearly lower than in the other cases due to dilution by the changing of $\psi_X$. However, the mean value of \pdx during periods of nonrotation of \pax (\first{} and \fourth{}) was $14\pm2\%$,  a factor of $1.4\pm0.2$ higher than the value of $10\pm1\%$ during the rotation (\second{} and \third{}). This result is similar to the findings of previous optical polarimetry studies: (1) \citet{2016MNRAS.457.2252B} observed that the ratio of \pdo of rotating to that of nonrotating cases is less than 1 in 18 out of the 27 observed rotations, and (2) \citet{2016MNRAS.462.4267J} reported a $26\%$ reduction in the average \pdo during rotational periods compared to nonrotational periods. Nonetheless, because of the limited amount of X-ray polarization data, this result should be verified with an increased number of measurements from future X-ray polarimetry monitoring observations.

The radio electric-vector position angle $\psi_{\mathrm{43GHz}}$ obtained from the 43~GHz Very Long Baseline Array (VLBA) observations (black line in Figure \ref{fig:polcontours}) differs from \pax and which also changes by $70\degr$ over the 7 months from \first\ to \fourth. The pronounced variation of \pax and $\psi_{\mathrm{43GHz}}$ contrasts with the steadier X-ray polarization observed in other sources of the same sub-class of blazars, whose synchrotron SED peaks at X-ray frequencies \citep[i.e., Mrk~501;][]{2022Natur.611..677L}. 

We have also investigated the spectral properties and X-ray activity of Mrk~421 through \swift\ monitoring observations. Figure \ref{fig:swift} presents the $\alpha$ and $\beta$ spectral parameters obtained from the \texttt{LOGPAR} model fit. We also derived the X-ray flux in the soft (0.3--2 keV), $F_{\mathrm{soft}}$, and hard (2--10 keV), $F_{\mathrm{hard}}$, energy bands, and define the hardness ratio as ($F_{\mathrm{hard}} - F_{\mathrm{soft}}$) / ($F_{\mathrm{hard}} + F_{\mathrm{soft}}$). We have found that the $\alpha$ parameter, representing the slope of the spectrum at the pivot energy, maintained similar values in \first{} and \fourth. However, during \second{} and \third, $\alpha$ decreased with time. In addition, the hardness ratio also varied during the \pax rotation, while its value in \first{} was consistent with that in \fourth. Nonetheless, in the case of \fourth, \swift{} observations were conducted only at the beginning and shortly after the \ixpe pointing, thus we cannot exclude the possibility of variations of spectral properties during \fourth.

The X-ray fluxes of Mrk~421 during all epochs of \ixpe\ observations were within $1\sigma$ of the median historical value \citep[e.g.,][]{2019ApJ...880...29L}. Nonetheless, we found significant fluctuations during \second{} and \third. Although the flux variations occurred during all of the \ixpe\ epochs (see, e.g., Figure \ref{fig:LC}), the discrepancy between the minimum and maximum counting rates was larger in \second{} and \third. Therefore, we suggest that the smooth \pax rotation behavior may accompany more pronounced spectral and flux fluctuations. 

\section{Multiwavelength polarization analysis}\label{sec:Multi}

\begin{figure*}[t!]
         \centering
         \includegraphics[width=\textwidth]{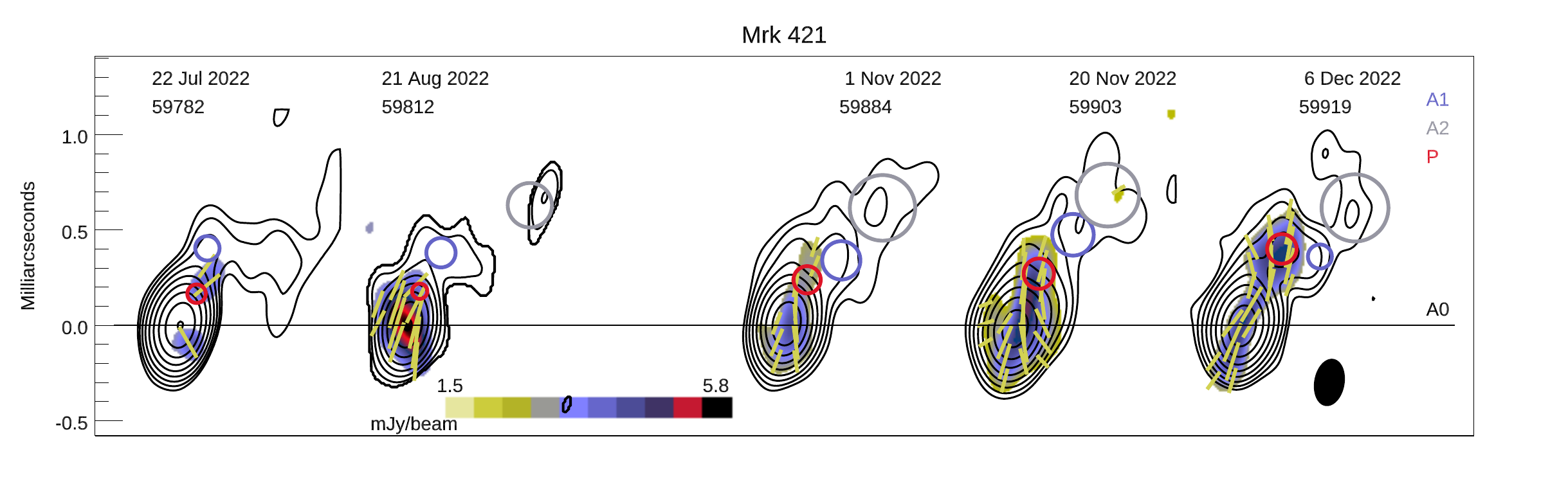}
\caption{VLBA total (contours) and polarized (color scale) intensity images of Mrk~421 at 43~GHz. The peak of the total intensity is 295 mJy/beam; yellow linear segments within the images indicate the direction of polarization; black horizontal line marks the position of the core, $A0$; colored circles indicate locations of jet features $A1$ (blue), $A2$ (gray), and $P$ (red). Images are convolved with the same elliptical Gaussian beam, which is shown in the bottom right corner by a black ellipse.}\label{fig:vlba}
\end{figure*}

Multiwavelength campaigns for the first three observations of Mrk~421 are reported in \citet{2022ApJ...938L...7D, DiGesu2023}. During \fourth, Mrk~421 was observed with the VLBA, the Effelsberg 100-m Radio Telescope, the Korean VLBI Network (KVN), the Submillimeter Array (SMA), the Kanata telescope, the Perkins telescope, and the Sierra Nevada Observatory (SNO, T90 telescope). Details of the observations and observatories can be found in Appendix \ref{app:multiobs}.

We have analyzed VLBA data obtained for Mrk~421 within the BEAM-ME (Blazars Entering the Astrophysical Multimessenger Era) program\footnote{\url{www.bu.edu/blazars/BEAM-ME.html}} during the period from MJD 59616 (2022 February 5) to MJD 59986 (2023 February 11) to investigate the parsec-scale jet behavior during the \ixpe observations. The data include total and polarized intensity images at 43~GHz at 12 epochs. The data reduction and modeling are described in \S\ref{app:vlbaobs}. The parsec-scale jet of Mrk~421 is strongly core-dominated, with the extended jet contributing an average of only $17\%$ of the core flux density at 43~GHz. We consider the core to be a stationary feature of the jet and designate it as $A0$. During the one-year period analyzed here, we have detected two stationary features in the jet in addition to the core, $A1$ and $A2$, located at distances of $\sim$0.4 and $\sim$0.8~mas from $A0$, respectively (see Table~\ref{tab:KnotParm}). Dominance by stationary structures is a well-documented property of the Mrk~421 jet \citep{Lico2012, J17, Lister2021, 2022ApJS..260...12W}. Nonetheless, we have found motion in the jet, which we identify with a prominent polarized feature, $P$, which has a subluminal apparent speed $\beta_{app}$=0.7$\pm$0.1 in units of $c$. Figure \ref{fig:vlba} presents a sequence of images at epochs where the knot is prominent. Figure~\ref{fig:1101_pol} shows that knot $P$ had a high degree of polarization, with $\psi_{\mathrm{P}}$ almost aligned with the jet direction, as expected if feature $P$ was a transverse shock in the jet. According to its kinematics, the knot passed through the core on 2022 March 14 (MJD 59653$\pm$80, Figure~\ref{fig:1101_move}). If one considers the sizes of $A0$ and $P$ listed in Table~\ref{tab:KnotParm}, plus the proper motion of $P$ (\S\ref{app:vlbaobs}), it should have taken 107$\pm$19 days for knot $P$ to leave the core completely, which should have occurred within 2022 June 10 -- July 8. This means that during \ixpe \first{}, \second{}, and \third{}, knot $P$ was within the millimeter-wave core.  

The Effelsberg 100-m Telescope observation was obtained within the framework of the QUIVER program (Monitoring the Stokes \textbf{Q}, \textbf{U}, \textbf{I} and \textbf{V} \textbf{E}mission of AGN jets in \textbf{R}adio) on 2022 December 2 (MJD~59915.11). The linear polarization was measured at 3 bands: 4.85\,GHz, 8.35\,GHz and 10.45\,GHz. Its polarization degree was found to be 0.9$\pm$0.2$\%$, 1.2$\pm$0.4$\%$, and 1.2$\pm$0.2$\%$. and its polarization angle 96.3$\pm$3.5$\degr$, 134.8$\pm$4.9$\degr$, and 159.2$\pm$7.2$\degr$, respectively. The KVN observations, with the antennas combined to act as a single dish, were performed on 2022 December 9 (MJD~59922.73) at 22, 43, 86, and 129~GHz. The linear polarization was measured as  $\Pi_{\mathrm{22GHz}}$=2.4$\pm$0.7$\%$ along $\psi_{\mathrm{22GHz}}$=30$\pm$12$\degr$ at 22 GHz and $\Pi_{\mathrm{43GHz}}$=2.7$\pm$0.5$\%$ along $\psi_{\mathrm{43GHz}}$=142$\pm$9$\degr$ at 43 GHz. The measured values of $\Pi$ at 86 GHz (3.1$\pm$2.3$\%$) and 129 GHz (4.4$\pm$2.0$\%$) are not significant detections. The polarization was measured with the SMA at 225.5~GHz as $\Pi_{\mathrm{225GHz}}$=2.0$\pm$0.3$\%$ along position angle $\psi_{\mathrm{225GHz}}$=163$\pm$3$\degr$. At the same time, the intrinsic polarization degree (after subtraction of the host-galaxy flux) in the optical R-band from SNO was $\Pi_O$=4.6$\pm$1.3$\%$ along $\psi_O$=206$\pm$9$\degr$. The Perkins telescope obtained optical R-band polarization covering \fourth{} from 2022 November 26 (MJD~59909) to December 17 (MJD~59930) to give the average optical polarization parameters just before and after \fourth, which can be compared with the SNO measurement. In the infrared J and optical R bands, the Kanata telescope data (not corrected for the host-galaxy depolarization) yield $\Pi_J$=2.41$\pm$0.02$\%$ and $\Pi_O$=2.1$\pm$0.03$\%$ along $\psi_J$=176$\pm$0.2$\degr$ and $\psi_O$=167.0$\pm$0.3$\degr$, respectively. The latter is similar to previous Mrk~421 R-band observations with the Kanata telescope, for which an intrinsic value of $\Pi_O$$\approx$3.3$\%$ was derived \citep{DiGesu2023}, which is consistent with the SNO measurement. All of the new multiwavelength polarimetry results conducted during \fourth{} are summarized in Table \ref{tab:multiwave}. 

Figure \ref{fig:multi_LC} exhibits the evolution of polarization properties of Mrk~421 obtained from the contemporaneous multiwavelength polarimetry campaign from just before \ixpe \first{} to shortly after \fourth. Throughout this time period, we find a similar strong chromatic behavior, with \pd 2--3 times higher at X-ray rather than at longer wavelengths. In contrast, the infrared, optical, and radio degrees of polarization were similar. Meanwhile, the polarization position angle exhibited marked changes, with different values during the various \ixpe pointings. The range of \pa observed in \first{} across millimeter to X-ray wavelengths was $\sim$30$\degr$, whereas a larger range of $\sim$90$\degr$ was evident in \fourth. Moreover, \citep{DiGesu2023} have reported that, during the rotation of \pax of \second{} and \third, \pa values at other wavelengths were consistent with each other, with a weak temporal variation. Hence, we conclude that the region where the polarized X-rays are emitted is mostly or completely distinct from that at longer wavelengths. 

Furthermore, Mrk~421 exhibited a clockwise \pao rotation of approximately 120$\degr$ over $\sim$40 days during the first three \ixpe observation periods. Similar behavior has been reported in previous optical polarimetry monitoring studies, where the polarization angle changed by $\sim$180$\degr$ over $\sim$50 days \citep{2015MNRAS.453.1669B, 2016MNRAS.462.4267J, 2016MNRAS.462.1775B, 2017ApJS..232....7F, 2018MNRAS.474.1296B}. The direction of the \pao rotation was opposite to the counter-clockwise \pax rotation observed by {\it IXPE} during \second{} and \third{}. 

\section{Discussion}\label{sec:DISCUSSION}

Previous \ixpe observations of HSPs Mrk~421 \citep{2022ApJ...938L...7D, DiGesu2023} and Mrk~501 \citep{2022Natur.611..677L} suggest that an energy-stratified shock can most readily explain the $\sim$2--3 times higher degree of polarization at X-ray rather than at longer wavelengths. In this work, we have found consistent multiwavelength measurements from \fourth. In this energy-stratified shock model, the relativistic electrons convert their energy to radiation as they move farther from the shock front \citep{Marscher1985,2018MNRAS.480.2872T}. The particles are efficiently accelerated at the shock front, where the magnetic field is relatively well ordered, and hence emit X-ray synchrotron radiation with relatively high polarization. Conversely, electrons lose energy as they propagate away from the shock, causing them to radiate at longer wavelengths, and the degree of polarization decreases as they encounter increasingly turbulent magnetic fields \citep{2022ApJ...938L...7D, DiGesu2023,2022Natur.611..677L}. Hence, higher \pd is predicted at higher frequencies. This effect ceases at wavelengths long enough that the electrons radiating at those wavelengths can cross the entire shocked region before losing most of their energy \citep{Marscher1985}. This limitation can explain the similar polarization from millimeter to optical wavelengths observed in Mrk~421. It is also possible that the longer wavelength emission occurred mainly in a relatively slow sheath surrounding a much faster X-ray emitting spine of the jet \citep[e.g.,][]{DiGesu2023}. This is consistent with the subluminal speed we have measured for knot $P$. In this case, the X-ray and longer wavelength polarization properties may not be related, consistent with their different position angles. 

We therefore consider two possible geometries for an energy-stratified jet in Mrk~421: linear and radial models, as suggested by \citet{2022ApJ...938L...7D, DiGesu2023}. In the case of linear geometry, the energy stratification and the \pa rotation can be explained by emission features propagating downstream in the jet, following the helical magnetic field. On the other hand, the radial structure can correspond to a helical, rotating innermost region and a surrounding layer, similar to the spine-sheath jet model \citep{2017MNRAS.466.3544C}. The currently available observational data are insufficient to discern between the linear and radial geometries. Nevertheless, the episodic variation in polarization observed during \fourth{} offers further insights into the internal geometry of the jet. For instance, it implies alternative perspectives on the geometric structure within the jet, including the intricate interactions between coexisting stable and rotating magnetic field structures, as well as stochastic transitions within the dominant magnetic field structure responsible for particle acceleration. Therefore, future observations of polarization variability are expected to yield further evidence about the geometric structure inside the jet.

On the other hand, despite the comparable multiwavelength results reported for Mrk~421 and Mrk~501, we have found a difference in the behavior of \pax between them. In the case of Mrk~421, $\psi_{\mathrm{43GHz}}$ and \pax exhibited significant variations without any consistent alignment with either each other or the direction of the jet axis. However, in the case of Mrk~501, an alignment was observed (within the uncertainties) between measurements of the position angle of the jet axis, $\psi_X$, and $\psi_{\mathrm{43GHz}}$ conducted within a month of each other \citep{2022Natur.611..677L}. Further \ixpe and VLBA observations of HSPs are needed to confirm whether, and if so, why, Mrk~421 is different in this regard.

We have found that the X-ray flux and hardness ratio of Mrk~421 were less variable during \first\ and \fourth, when \pax was essentially constant, than during the \pax rotation of \second\ and \third. This implies different physical conditions within the jet between the rotating and nonrotating states. Instead, the similar values of \pdx across all observations suggest that the basic particle acceleration scenario remained roughly independent of the magnetic field geometry. This can be accommodated within the energy-stratified shock scenario, since the degree of order of the magnetic field could be similar whether the shock moves along a straight or helical trajectory. The flux and spectral variations during the rotation event could have been caused by changes in the Doppler factor as the shock executed helical motion, although the data are too sparsely sampled to test this. Future X-ray spectroscopic and polarimetric observations with improved time resolution can potentially test whether spiral motion along a helical magnetic field causes cyclical Doppler factor variations that lead to observed variations in flux, hardness ratio, and polarization.

Our examination of a sequence of 43 GHz VLBA images has revealed the presence of a prominent, highly polarized knot moving away from the ``core'' at a speed of $0.7c$ during a time span that includes \first\ - \third\
(see Figure~\ref{fig:1101_move}).

This finding suggests a potential connection between the morphological changes near the jet core region and the variability in polarization, as discussed in \citet{DiGesu2023}. The knot could represent a shock containing relativistic electrons accelerated up to Lorentz factor $\sim$10$^6$ that radiate at X-ray energies. However, in this scenario, it is difficult to explain the difference in the behavior of $\psi_{\mathrm{X}}$ between  \first\ and \second\ - \third, unless the geometry of the magnetic field varies across the core, with a tighter helical structure toward the downstream end. However, our analysis does not indicate a direct connection between the X-ray polarization position angle and that of either the core or knot $P$ (see Figure \ref{fig:1101_pol}), although the degree of polarization of $P$, 15-20\%, is comparable to \pdx. This apparent lack of connection supports the conclusion, drawn above, that emission regions at longer wavelengths (millimeter, IR, and optical) were separate from, or only partially coincident with, that of the X-ray emission.     

In the case of \fourth, there is no apparent connection between the values of $\psi$ of the polarized features in the jet observed on 2022 December 6 (MJD 59919) and \pax (Fig.~\ref{fig:1101_pol}). However, Figure~\ref{fig:1101_lc} indicates an increase of the core flux density in 2023 February (MJD 59986) that could be a signature of an emerging moving feature that would have been upstream of the 43 GHz VLBI core during \ixpe\ \fourth. Further combined \ixpe and VLBI monitoring could help to clarify whether there is any relation between the X-ray emission regions and features seen in the jet at millimeter wavelength.

\begin{figure}[t!]
\centering
\includegraphics[width=0.5\textwidth]{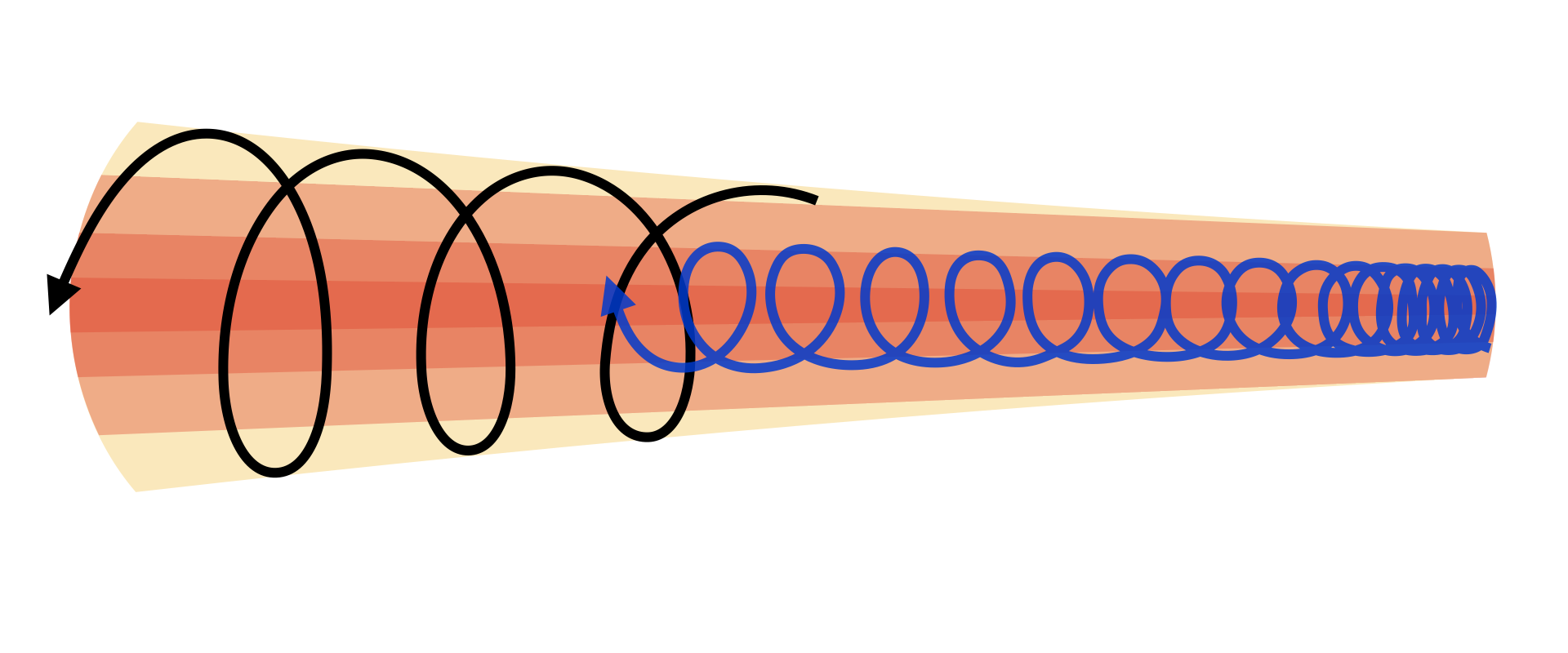}
\caption{Schematic diagram of double helical magnetic field components inside the jet. The downstream direction of the jet is to the left. The arrows indicate each helical magnetic field component involved in the emission at different wavelengths (blue: X-ray, black: longer wavelengths).}\label{fig:model}
\end{figure}

We note that, from \first\ to \third, the polarization angle at optical, IR, and radio wavelengths rotated in the opposite direction relative to the 5-day rotation of \pax during \second\ and \third. This finding supports the conclusion that the X-ray emission region is separate from that at longer wavelengths. The observed similarity in radio to optical polarization properties implies that the emission at these wavelengths originates from spatially interconnected regions. The longer (relative to X-ray) rotation time scale suggests that the X-ray emission region is smaller than that of these other wavelengths.
Although the polarization vector rotation at longer wavelengths could be explained by propagation of a larger emission feature down a helical magnetic field (as in Figure \ref{fig:model}), the long-term behavior of the optical polarization of Mrk~421 implies stochastic, rather than systematic, variations of both $\psi_O$ and $\Pi_O$ \citep{Marscher2022}.

\section{Conclusions}\label{sec:CONCLUSIONS}

We have reported X-ray, optical, IR, and radio polarization and flux measurements of the HSP blazar Mrk~421, including four \ixpe pointings between 2022 May and December. Such observations probe the magnetic field structure and particle acceleration mechanisms inside the jets of blazars. The combined 7 months of observations sample the time and energy dependence of the X-ray polarization. Over this time span, \pax varied over the full range of $\sim180\degr$, including a 5-day episode of rotation, while the degree of polarization maintained a value between $\sim$10--15$\%$ across all \ixpe pointings. The X-ray flux varied by a higher fraction during the rotation than during the two \ixpe observations without rotations. 

The simultaneous multiwavelength polarimetry results over four \ixpe observations provide evidence useful for constraining the physics of the jet. The degree of X-ray polarization was typically $\sim$2--3 times greater than that at longer wavelengths at all epochs sampled, while the polarization angles fluctuated. The discrepancy between the results of X-ray compared with radio, IR, and optical polarimetry supports the previous conclusion that the X-ray emission region is distinct from that at longer wavelengths in HSP blazars \citep{2022Natur.611..677L,2022ApJ...938L...7D,DiGesu2023}. As with these previous studies, we conclude that the observations are consistent with the energy stratified shock model, with the level of turbulence increasing with distance from the shock front.

One difference between Mrk~421 and Mrk~501 is that there is no apparent correlation between the direction of the jet from VLBA and $\psi_X$ in the former. While this could be due to the bending of the jet from the X-ray to the radio emitting region, amplified by the narrow angle to the line of sight, the optical polarization angle is also much more highly variable in Mrk~421 than in Mrk~501 \citep{Marscher2022}. This implies an intrinsic difference between the two objects that should be explored with further observations.

Following \citet{DiGesu2023}, we have proposed a linear and a radial stratification scenarios to explain the rotation behavior of $\psi_{X}$. The accompanying spectral variation during the \pax rotation suggests that the physical conditions of the jet, such as the energy distribution of relativistic electrons, differed between the periods of rotation and nonrotation. In addition, we reported rotation in the opposite direction of the \pa between the X-ray and other wavelengths, with the latter occurring over a much longer time scale. This could potentially be interpreted by the presence of multi helical magnetic field structures inside the jet. Morphological changes in the parsec-scale jet, possibly associated with the contemporaneous emergence of a new knot of emission observed to move down the jet in 43 GHz VLBA images, may be linked to the \pax rotation, although differences in the radio and X-ray polarization angles argue against such a connection.

In conclusion, the present study continues to develop a new perspective on the physical and geometrical features of the magnetic field inside the jets of blazars by employing polarimetry that extends from radio to X-ray wavelengths. The \ixpe observations, incorporating data at other wavelengths, have played a significant role in constraining the emission arising from the innermost regions of the jet. The polarization properties, sampled over different time scales and energy regimes, suggest a possible connection between spectral and polarization variations. The connection may include morphological changes in radio images, which coincided with a period of \pax rotation. However, due to infrequent data sampling, there remain uncertainties regarding apparent  correlations, which can be chance coincidences. In addition, thus far the \ixpe observations of Mrk~421 have been obtained when the blazar was in average activity states. It is of great interest to determine whether the polarization and physical properties change during strong flaring events. As the \ixpe mission continues, further studies of Mrk~421 and other blazars are expected to provide the data needed to improve our understanding of the magnetic field geometry and particle acceleration processes in relativistic jets.

\begin{acknowledgements} 
The authors thank the anonymous referee for comments that improved this manuscript. The Imaging X-ray Polarimetry Explorer (IXPE) is a joint US and Italian mission.  The US contribution is supported by the National Aeronautics and Space Administration (NASA) and led and managed by its Marshall Space Flight Center (MSFC), with industry partner Ball Aerospace (contract NNM15AA18C).  The Italian contribution is supported by the Italian Space Agency (Agenzia Spaziale Italiana, ASI) through contract ASI-OHBI-2017-12-I.0, agreements ASI-INAF-2017-12-H0 and ASI-INFN-2017.13-H0, and its Space Science Data Center (SSDC), and by the Istituto Nazionale di Astrofisica (INAF) and the Istituto Nazionale di Fisica Nucleare (INFN) in Italy. This research used data products provided by the IXPE Team (MSFC, SSDC, INAF, and INFN) and distributed with additional software tools by the High-Energy Astrophysics Science Archive Research Center (HEASARC), at NASA Goddard Space Flight Center (GSFC). The IAA-CSIC group acknowledges financial support from the grant CEX2021-001131-S funded by MCIN/AEI/10.13039/501100011033 to the Instituto de Astrof\'isica de Andaluc\'ia-CSIC and through grant PID2019-107847RB-C44. The QUIVER data are based on observations with the 100-m telescope of the MPIfR (Max-Planck-Institut f\"ur Radioastronomie) at Effelsberg. Observations with the 100-m radio telescope at Effelsberg have received funding from the European Union’s Horizon 2020 research and innovation programme under grant agreement No 101004719 (ORP). The POLAMI observations were carried out at the IRAM 30m Telescope. IRAM is supported by INSU/CNRS (France), MPG (Germany), and IGN (Spain). Some of the data are based on observations collected at the Observatorio de Sierra Nevada, owned and operated by the Instituto de Astrof\'{i}sica de Andaluc\'{i}a (IAA-CSIC). Further data are based on observations collected at the Centro Astron\'{o}mico Hispano en Andalucía (CAHA), operated jointly by Junta de Andaluc\'{i}a and Consejo Superior de Investigaciones Cient\'{i}ficas (IAA-CSIC). The Submillimetre Array is a joint project between the Smithsonian Astrophysical Observatory and the Academia Sinica Institute of Astronomy and Astrophysics and is funded by the Smithsonian Institution and the Academia Sinica. Mauna Kea, the location of the SMA, is a culturally important site for the indigenous Hawaiian people; we are privileged to study the cosmos from its summit. Some of the data reported here are based on observations made with the Nordic Optical Telescope, owned in collaboration with the University of Turku and Aarhus University, and operated jointly by Aarhus University, the University of Turku, and the University of Oslo, representing Denmark, Finland, and Norway, the University of Iceland and Stockholm University at the Observatorio del Roque de los Muchachos, La Palma, Spain, of the Instituto de Astrofisica de Canarias. E. L. was supported by Academy of Finland projects 317636 and 320045. The data presented here were obtained [in part] with ALFOSC, which is provided by the Instituto de Astrofisica de Andalucia (IAA) under a joint agreement with the University of Copenhagen and NOT. We acknowledge funding to support our NOT observations from the Finnish Centre for Astronomy with ESO (FINCA), University of Turku, Finland (Academy of Finland grant nr 306531). We are grateful to Vittorio Braga, Matteo Monelli, and Manuel S\"{a}nchez Benavente for performing the observations at the Nordic Optical Telescope.  Part of the French contributions is supported by the Scientific Research National Center (CNRS) and the French spatial agency (CNES). The research at Boston University was supported in part by National Science Foundation grant AST-2108622, NASA Fermi Guest Investigator grants 80NSSC21K1917 and 80NSSC22K1571, and NASA Swift Guest Investigator grant 80NSSC22K0537. This study was based in part on observations conducted using the Perkins Telescope Observatory (PTO) in Arizona, USA, which is owned and operated by Boston University. This research was conducted in part using the Mimir instrument, jointly developed at Boston University and Lowell Observatory and supported by NASA, NSF, and the W.M. Keck Foundation. We thank D.\ Clemens for guidance in the analysis of the Mimir data. This work was supported by JST, the establishment of university fellowships toward the creation of science and technology innovation, Grant Number JPMJFS2129. This work was supported by Japan Society for the Promotion of Science (JSPS) KAKENHI Grant Numbers JP21H01137. This work was also partially supported by the Optical and Near-Infrared Astronomy Inter-University Cooperation Program from the Ministry of Education, Culture, Sports, Science and Technology (MEXT) of Japan. We are grateful to the observation and operating members of the Kanata Telescope.  This research has made use of data from the RoboPol program, a collaboration between Caltech, the University of Crete, IA-FORTH, IUCAA, the MPIfR, and the Nicolaus Copernicus University, which was conducted at Skinakas Observatory in Crete, Greece. D.B., S.K., R.S., N. M., acknowledge support from the European Research Council (ERC) under the European Unions Horizon 2020 research and innovation program under grant agreement No.~771282. CC acknowledges support from the European Research Council (ERC) under the HORIZON ERC Grants 2021 program under grant agreement No. 101040021. We acknowledge the use of public data from the Swift data archive. Based on observations obtained with XMM-Newton, an ESA science mission with instruments and contributions directly funded by ESA Member States and NASA. The Very Long Baseline Array is an instrument of the National Radio Astronomy Observatory. The National Radio Astronomy Observatory is a facility of the National Science Foundation operated under a cooperative agreement by Associated Universities, Inc. S. Kang, S.-S. Lee, W. Y. Cheong, S.-H. Kim, and H.-W. Jeong  were supported by the National Research Foundation of Korea (NRF) grant funded by the Korea government (MIST) (2020R1A2C2009003). The KVN is a facility operated by the Korea Astronomy and Space Science Institute. The KVN operations are supported by KREONET (Korea Research Environment Open NETwork) which is managed and operated by KISTI (Korea Institute of Science and Technology Information). The VLBA is an instrument of the National Radio Astronomy Observatory. The National Radio Astronomy Observatory is a facility of the National Science Foundation operated under cooperative agreement by Associated Universities, Inc.

\end{acknowledgements}

\bibliographystyle{aa_url.bst}
\bibliography{references.bib}

\begin{appendix} 
\section{Observation and data reduction}\label{sec:Appendix}

\subsection{IXPE data}

We obtained cleaned level 2 \ixpe data processed by a standard \ixpe pipeline from the Science Operation Center (SOC)\footnote{\url{https://heasarc.gsfc.nasa.gov/docs/ixpe/analysis/IXPE-SOC-DOC-009-UserGuide-Software.pdf}}. The pipeline includes  the photoelectron events correction in the Gas Pixel Detector \citep[GPD,][]{2001Natur.411..662C, 2007NIMPA.579..853B, 2012AdSpR..49..143F, 2021APh...13302628B}, as well as the photo-electron track reconstruction process based on standard moments analysis \citep{2003SPIE.4843..372B, fabiani2014astronomical, 2022AJ....163..170D}. In particular, the pipeline rectifies the fluctuations in gain properties affected by gas conditions (e.g., temperature and pressure), nonuniform charging of the Gas Electron Multiplier (GEM) material, and polarization artifacts induced by spurious modulation \citep{Rankin_2022}. The level 2 data of \ixpe contain the time of arrival, position, and energy of each photon event, along with polarization information represented by the \Q and \U Stokes parameters.

The scientific data analysis was performed using the public \ixpeobssim software version 30.2.1 \citep{2022SoftX..1901194B}. First, we extracted the source and background data using an optimized region selection criterion as suggested by \citet{2023AJ....165..143D}. The source data were extracted from a circular region with a 60\arcsec radius, while the background data were extracted from an annular source-free region with inner and outer radii of 150\arcsec and 300\arcsec, respectively. Both regions were centered on the source position in the detector frame. Next, we created the polarization cube (\texttt{PCUBE}) and the Stokes parameter spectra (\I, \Q, and \U) using \texttt{xpbin}. To eliminate the influence due to the background, we applied the background subtraction technique \citep[][]{2022SoftX..1901194B} and created \texttt{PCUBE}s of the source and background for each detector unit. The final source polarization properties were recalibrated by considering the \texttt{BACKSCALE} ratio between the source and background ($\sim0.05$). In a similar way, we produced three Stokes parameter spectra for three detectors, totaling 9 spectra, using the \texttt{PHA1}, \texttt{PHAQ}, and \texttt{PHAU} algorithms in \texttt{xpbin}. We grouped the \I spectra with a minimum of 30 counts in each energy bin, as required for the $\chi^2$ statistics in the fits, and applied constant energy binning for the \Q and \U spectra for every 0.2 keV interval. We employed the recent version of the \ixpe calibration database files (CALDB 20221020) contained in both \ixpeobssim and the \textsc{heasoft} package (v6.31.1) for both the \texttt{PCUBE} and the Stokes parameter spectra. In particular, we utilized the weighted analysis method \citep{2022AJ....163..170D} on the \I, \Q, and \U spectra to improve the significance of our measurements with \texttt{alpha075} response matrices. The flux variability was analyzed using a light curve created with the \texttt{LC} algorithms in \texttt{xpbin} of \texttt{ixpeobssim}.

The polarization error contour (Figure \ref{fig:ixpe_polcontour}) was drawn from calibrated normalized ${\mathcal Q} \,(=\Q/\I)$ and ${\mathcal U} \,(=\U/\I)$ at specific confidence levels (68.27\%, 90.00\%, and 99.0\%) according to: 
\begin{equation} \label{eq:polar}
\begin{split}
&{\mathcal Q}_{C.L.} = {\mathcal Q}_{0} + \epsilon \cos(\zeta),\\ 
&{\mathcal U}_{C.L.} = {\mathcal U}_{0} + \epsilon \sin(\zeta),\\
\end{split}
\end{equation}
\noindent where ${\mathcal Q}_0$ and ${\mathcal U}_0$ represent the averaged Stokes parameter values, and ${\mathcal Q}_{C.L.}$ and ${\mathcal U}_{C.L.}$ denote the ${\mathcal Q}$ and ${\mathcal U}$ values of a given confidence level calculated with two degrees of freedom, respectively. We note that, since we are considering two dependent variables, $\Pi$ and $\psi$, at the same time, the error should be recalculated based on the $\chi^2$ distribution with 2 degrees of freedom ({\small $\epsilon = \sigma${\tiny$\sqrt{\chi^2_{\mathrm {2 \, d.o.f.}}}$}, $\sigma= \sigma_{\mathcal Q} = \sigma_{\mathcal U}$}). The variable $\zeta$ follows the angle distribution from 0 to $2 \pi$.

\subsection{Spectroscopic X-ray data}

During \ixpe \fourth, a snapshot of about 5 ks was performed with \xmm{} in timing mode to limit photon pile-up. To extract the science products, we used the \xmm{} Science Analysis Software (SAS), version 21, with the most updated CCF (current calibration files). Two boxed regions were adopted to extract the source spectra and background. In particular, we chose the source box to have a width of 27 pixels, as this size was found to maximize the signal-to-noise ratio. (Details of this procedure can be found in \citealt{Piconcelli2004}.) Then, the resulting spectrum was binned to achieve at least 30 counts in each energy bin. To allow simultaneous spectropolarimetric fitting within \texttt{XSPEC}, the keyword ``XFLT0001 Stokes:0'' was added to the headers of the corresponding \texttt{PHA} files.

The \swift{} monitoring consisted of $\sim$1 ks long observations performed in Windowed Timing (WT) mode. Raw data were reduced using the standard commands within the \textit{XRT} Data Analysis Software (XRTDAS, v. 3.6.1) and adopting the latest calibration files available in the \swift{} CALDB (version 20210915). The cleaned event files were used to extract the source spectrum from a circular region with a radius of 47$\arcsec$, while the background was extracted using a blank sky WT observation available in the Swift archive, also within a circular region of radius 47$\arcsec$. The resulting spectra were subsequently binned, requiring at least 25 counts in each energy bin.

\begin{figure}[t]
\centering         
\includegraphics[width=1.\columnwidth]{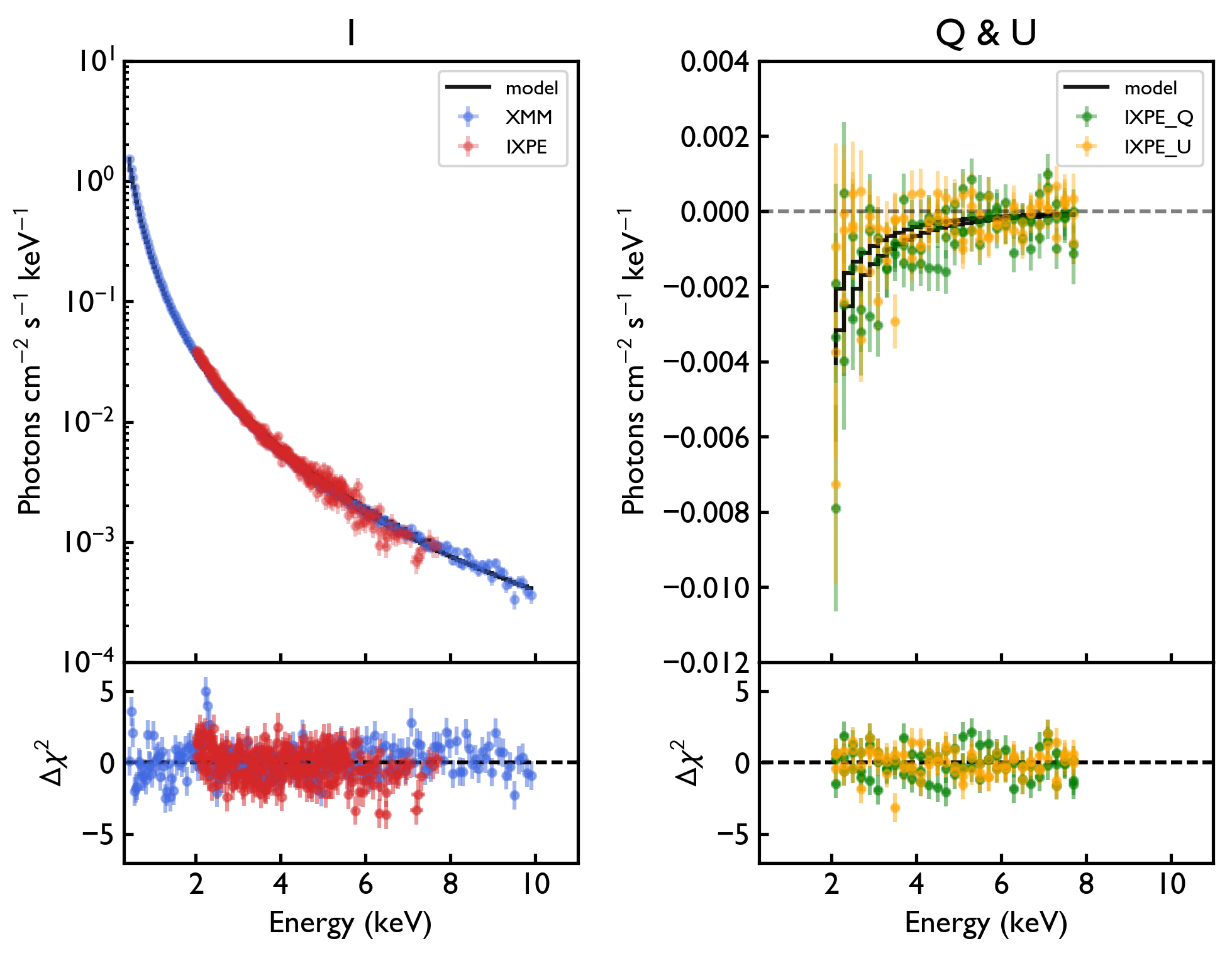}\\
\caption{Spectropolarimetric fit of {\it XMM-Newton} (blue) and \ixpe \I (red), \Q (green), and \U (orange) spectra, with residuals of the best-fit model. The left panel displays the \I spectrum with a log-parabolic model \citep{2004A&A...413..489M}, while the right panel presents the constant polarization model fit with the \texttt{POLCONST} model for \Q and \U spectra. The black line in both panels indicates the model fit result.
}\label{fig:xspec}
\end{figure}

\begin{table}[t]
\caption{Best-fit parameters from our simultaneous spectropolarimetric analysis. \label{tab:XSPEC}}\centering
\begin{tabular}{lrcr}
\hline\hline
\noalign{\smallskip}
\multicolumn{4}{c}{\texttt{MODEL} = \texttt{CONSTANT} $\times$ \texttt{TBABS} $\times$ \texttt{LOGPAR} $\times$ \texttt{POLCONST}}\\
\hline\hline
\noalign{\smallskip}
Component & Parameter & Units & Value ($\pm 1 \sigma$)\\
\hline
\noalign{\smallskip}
\texttt{LOGPAR} & $\alpha$ & & 2.82 $\pm$0.005\\
 & $\beta$ & & 0.20 $\pm$0.004\\
 & $pivot \;E$ & $\rm{keV}$ & 5. (\textcolor{magenta}{f})\\
 & $Norm$ & & 2.96e-3 $\pm$1.e-5\\
 \noalign{\smallskip}
\hline
\noalign{\smallskip}
\texttt{POLCONST} & \pdx & ($\%$) & 14 $\pm$1 \\
				  & \pax & ($\degr$) & 107 $\pm$3\\
\noalign{\smallskip}
\hline
\noalign{\smallskip}
\texttt{TBABS} & $N_H$ & $10^{22} {\rm cm^{-2}}$ & 1.34e-2 (\textcolor{magenta}{f})\\
\noalign{\smallskip}
\hline
\noalign{\smallskip}
\texttt{CONSTANT} & {\it XMM-Newton} &  & 1. (\textcolor{magenta}{f})\\
 & \ixpe \; DU 1&  & 1.11 $\pm$ 0.005\\
 & \ixpe \; DU 2&  & 1.06 $\pm$ 0.005\\
 & \ixpe \; DU 3&  & 1.01 $\pm$ 0.005\\
\noalign{\smallskip}
\hline
\noalign{\smallskip}
\multicolumn{4}{r}{$\chi^2$ / ${\rm d.o.f.}$= 740/650}\\
\noalign{\smallskip}
\hline
\end{tabular}
\tablefoot{Errors are $1\sigma$ confidence and (\textcolor{magenta}{f}) denotes a fixed parameter.}
\end{table}

\subsection{Radio, Infrared and optical data}\label{app:multiobs}

During all of the IXPE pointings, we observed Mrk~421 at millimeter (radio), infrared (IR), and optical wavelengths, measuring both flux density and linear polarization. The observations and data analysis for the first three observations can be found in detail in \citet{2022ApJ...938L...7D, DiGesu2023}; here we provide only a short description. For \fourth, radio, IR, and optical observations were provided by QUIVER at the Effelsberg telescope, KVN, SMA \citep{Ho2004}, Hiroshima Optical and Near-InfraRed camera (HONIR, \citealp{Akitaya2014}) at the Kanata telescope, the Perkins telescope, and T90 at the Sierra Nevada Observatory (SNO). 

The QUIVER observations were performed at several radio bands (depending on receiver availability and weather conditions) from 2.6 GHz to 44 GHz (11 cm to 7 mm wavelength) using six receivers located at the secondary focus of the 100-m Effelsberg Radio Telescope (S110mm, S45mm, S28mm, S20mm, S14mm, and S7mm). The receivers are equipped with two orthogonally polarized feeds (either circular or linear) that can deliver polarimetric parameters using either conventional polarimeters or by connecting the SpecPol spectropolarimetric backend. Instrumental polarization is calibrated via observations of both polarized and unpolarized calibrators performed in each session, and then removed from the data \citep[e.g.,][]{2018A&A...609A..68M, 2003A&A...401..161K}. The polarized intensity, degree, and position angle were derived from the Stokes I, Q, and U cross-scans. The total flux density was successfully recovered at 13 bands between 4.85\,GHz and 43.75\,GHz. The calibrators 3C\,286, 3C\,48, and NGC\,7027 were used for the total flux and polarization calibration \citep[e.g.,][]{2018A&A...609A..68M, 2003A&A...401..161K}. 

Observations were conducted with the KVN simultaneously at 4 frequencies from 22 to 129~GHz in single-dish mode. This was achieved by using the Tamna (22, 43~GHz) and Yonsei (86,129~GHz) antennas with circularly polarized feed horns to conduct two-frequency dual-polarization observations. The polarization angle was calibrated using the Crab nebula (152$^\circ$; \citealp{Aumont2010}), and the polarization degree using Jupiter (unpolarized) and 3C286 (polarized, \citealp{Agudo2012}) following \cite{Kang2015}. 

The SMA observations were taken at 225.538~GHz (1.3~mm) through the SMAPOL program (SMA Monitoring of AGN with POLarization). The SMAPOL observations were taken on 2022 December 7 (MJD~59920) in full polarization mode \citep{Marrone2008} using the SWARM correlator \citep{Primiani2016}, and calibrated with the MIR software package \footnote{\url{https://lweb.cfa.harvard.edu/~cqi/mircook.html}}.

Observations using HONIR were taken on 2022 December 6 (MJD~59919.7537) in the R and J bands\footnote{\url{http://hasc.hiroshima-u.ac.jp/instruments/honir/filters-e.html}}. Each observation comprised four exposures in different positions of the half-wave plate \citep{Kawabata1999}, which were then used to estimate the Stokes parameters and calculate the polarization degree and angle. The host galaxy of Mrk~421 contributes a significant fraction of unpolarized flux to the total emission, thereby reducing the polarization. None of the polarization degree estimates from Kanata have been corrected for the host-galaxy contribution due to the lack of the photometry data needed for such a correction, and hence they should be considered as lower limits to the intrinsic polarization degree of the blazar component.

The optical photometric and polarimetric observations were also obtained at the 1.8~m Perkins telescope (Flagstaff, AZ, USA) with the PRISM camera\footnote{\url{https://www.bu.edu/prism/}} in R band performed before and after \fourth. The camera includes a polarimeter with a rotating half-wave plate. Each polarization observation has consisted of four consecutive exposures of 60~s at instrumental position angles 0$\degr$, 90$\degr$, 45$\degr$, and 135$\degr$ of the waveplate to calculate the normalized Stokes parameters \Q and \textit{U}. 

The SNO observations were performed using polarized filters oriented to represent different positions of a half-wave plate, similar to a conventional polarimeter. The data are then analyzed following standard photometric procedures. Mrk~421 was observed on 2022 December 7 in the R band. In this case, following \citet{Nilsson2007} and \citet{Hovatta2016}, we have corrected the host-galaxy contribution. Several observations were taken within the same night. We report the weighted average and uncertainty after considering all of the intra-night observations to account for variations of the system and the observing conditions during the rotation of the filter wheel. 

\begin{table*}[t!]
\caption{Multiwavelength polarization properties of Mrk~421 \label{tab:multiwave}}\centering
\begin{tabular}{lllll}
\hline\hline           
\noalign{\smallskip}
Telescope & Band & Dates & Radio Polarization & Radio Flux Density\\
& (GHz) & (YYYY-MM-DD) & $\Pi_{\rm R}$(\%) \qquad $\psi_{\rm R}$ ($\degr$)  & ($10^{-25}$\ergsch{})\\
\hline
\noalign{\smallskip}
KVN & 22 & 2022-12-09 & 2.4 $\pm$ 0.7 \quad 210 $\pm$ 12 & 45 $\pm$ 4 \\
KVN & 43 & 2022-12-09 & 2.7 $\pm$ 0.5 \quad 142 $\pm$ 9  & 40 $\pm$ 2 \\
KVN & 86 & 2022-12-09 & 3.1 $\pm$ 2.3 \quad 153 $\pm$ 15 & 34 $\pm$ 7 \\
KVN & 129 & 2022-12-09 & 4.4 $\pm$ 2.0 \quad 191 $\pm$ 16 & 71 $\pm$ 19\\
SMA & 230 & 2022-12-07 & 2.0 $\pm$ 0.3 \quad 163 $\pm$ 3  & 20 $\pm$ 2 \\ 
Effelsberg & 4.85 & 2022-12-02 & 0.9 $\pm$ 0.2 \quad 96 $\pm$ 4  & 59.4 $\pm$ 1.0 \\
Effelsberg & 8.35 & 2022-12-02 & 1.2 $\pm$ 0.4 \quad 135 $\pm$ 5  & 50.1 $\pm$ 0.5 \\
Effelsberg & 10.45 & 2022-12-02 & 1.2 $\pm$ 0.2 \quad 159 $\pm$ 7  & 48.7 $\pm$ 1.1 \\ 
VLBA & 43 & 2022-12-06 & 1.7 $\pm$ 0.6  \quad 171 $\pm$ 10 & 25.9 $\pm$ 3.3\\
VLBA & 43 & 2022-11-20 & 2.1 $\pm$ 0.4  \quad 178 $\pm$ 7 & 22.9 $\pm$ 3.3\\
VLBA & 43 & 2022-11-01 & 1.5 $\pm$ 0.3  \quad 181  $\pm$ 8 & 27.5 $\pm$ 3.5\\
VLBA & 43 & 2022-08-21 & 2.0 $\pm$ 0.6  \quad 166 $\pm$ 9 & 29.4 $\pm$ 3.6\\
VLBA & 43 & 2022-07-22 & 1.3 $\pm$ 0.5  \quad 215 $\pm$ 13 & 28.8 $\pm$ 3.6\\
VLBA & 43 & 2022-07-15 & < 1.1\quad  \qquad Undetermined & 29.0 $\pm$ 4.1\\
\noalign{\smallskip}
\hline\hline
\noalign{\smallskip}
Telescope & Band & Dates & Infrared Polarization & Infrared flux density\\
& ($\mu m$) & (YYYY-MM-DD) & $\Pi_{\rm IR}$(\%) \qquad $\psi_{\rm IR}$ ($\degr$) & ($10^{-25}$\ergsch{}) \\
\hline
\noalign{\smallskip}
Kanata \footnotemark[1] & J:1.1 & 2022-12-07 & 2.1 $\pm$ 0.03 \quad 167.0 $\pm$ 0.3  & 1.65 $\pm$ 0.03  \\ 
\noalign{\smallskip}
 \hline\hline
\noalign{\smallskip}
 
Telescope & Band & Dates & Optical Polarization & Optical flux density\\
& (nm) & (YYYY-MM-DD)  & $\Pi_{\rm O}$(\%) \qquad $\psi_{\rm O}$ ($\degr$)  & ($10^{-25}$\ergsch{})\\
\noalign{\smallskip}
\hline
\noalign{\smallskip}
SNO & R:653 & 2022-12-07 & 4.6 $\pm$ 1.3 \qquad 206 $\pm$ 9 & 2.11 $\pm$ 0.03 \\ 
Kanata \footnotemark[1] & R:653 & 2022-12-07 & 2.41 $\pm$ 0.02 \quad 176 $\pm$ 0.2 & -- \\
Perkins& R:653 & 2022-12-17 & 2.65 $\pm$ 0.21 \quad 151.9 $\pm$ 2.0 & 2.75 $\pm$ 0.03\\
Perkins& R:653 & 2022-12-17 & 2.68 $\pm$ 0.21 \quad 156.2 $\pm$ 2.3 & 2.74 $\pm$ 0.03\\
Perkins& R:653 & 2022-12-17 & 2.51 $\pm$ 0.21 \quad 155.9 $\pm$ 2.5 & 2.75 $\pm$ 0.03\\
Perkins& R:653 & 2022-12-16 & 1.12 $\pm$ 0.27 \quad 156.2 $\pm$ 7.2 & 2.87 $\pm$ 0.03\\
Perkins& R:653 & 2022-12-15 & 1.48 $\pm$ 0.28 \quad 160.6 $\pm$ 5.7 & 2.77 $\pm$ 0.03\\
Perkins& R:653 & 2022-12-14 & 1.74 $\pm$ 0.30 \quad 125.1 $\pm$ 5.3 & 2.72 $\pm$ 0.02\\
Perkins& R:653 & 2022-12-01 & 6.89 $\pm$ 0.39 \quad 161.7 $\pm$ 1.8 & 2.98 $\pm$ 0.04\\
Perkins& R:653 & 2022-12-01 & 7.11 $\pm$ 0.37 \quad 164.5 $\pm$ 1.7 & 2.99 $\pm$ 0.03\\
Perkins& R:653 & 2022-12-01 & 6.91 $\pm$ 0.37 \quad 159.0 $\pm$ 1.6 & 3.00 $\pm$ 0.03\\
Perkins& R:653 & 2022-11-28 & 4.76 $\pm$ 0.25 \quad 179.1 $\pm$ 1.6 & 2.99 $\pm$ 0.03\\
Perkins& R:653 & 2022-11-28 & 4.39 $\pm$ 0.28 \quad 164.4 $\pm$ 1.9 & 3.00 $\pm$ 0.04\\
Perkins& R:653 & 2022-11-28 & 5.22 $\pm$ 0.22 \quad 164.7 $\pm$ 1.3 & 3.00 $\pm$ 0.03\\
Perkins& R:653 & 2022-11-28 & 4.22 $\pm$ 0.22 \quad 168.6 $\pm$ 1.6 & 2.99 $\pm$ 0.03\\
Perkins& R:653 & 2022-11-27 & 5.79 $\pm$ 0.26 \quad 162.3 $\pm$ 1.3 & 2.98 $\pm$ 0.03\\
Perkins& R:653 & 2022-11-27 & 5.79 $\pm$ 0.21 \quad 165.3 $\pm$ 0.8 & 2.99 $\pm$ 0.02\\
Perkins& R:653 & 2022-11-27 & 5.02 $\pm$ 0.21 \quad 163.7 $\pm$ 0.8 & 2.98 $\pm$ 0.02\\
Perkins& R:653 & 2022-11-27 & 5.29 $\pm$ 0.21 \quad 163.3 $\pm$ 0.4 & 2.99 $\pm$ 0.02\\
Perkins& R:653 & 2022-11-26 & 6.42 $\pm$ 0.21 \quad 151.5 $\pm$ 0.8 & 2.86 $\pm$ 0.02\\
Perkins& R:653 & 2022-11-26 & 5.71 $\pm$ 0.23 \quad 152.0 $\pm$ 1.1 & 2.87 $\pm$ 0.02\\
Perkins& R:653 & 2022-11-26 & 5.44 $\pm$ 0.21 \quad 155.0 $\pm$ 0.9 & 2.86 $\pm$ 0.02\\

\noalign{\smallskip}
\hline
\end{tabular}
\tablefoot{ \footnotemark[1]Not corrected for dilution of polarization by unpolarized starlight from the host galaxy.}
\end{table*}

\begin{figure*}[t!]
\centering         
\includegraphics[width=1.\textwidth]{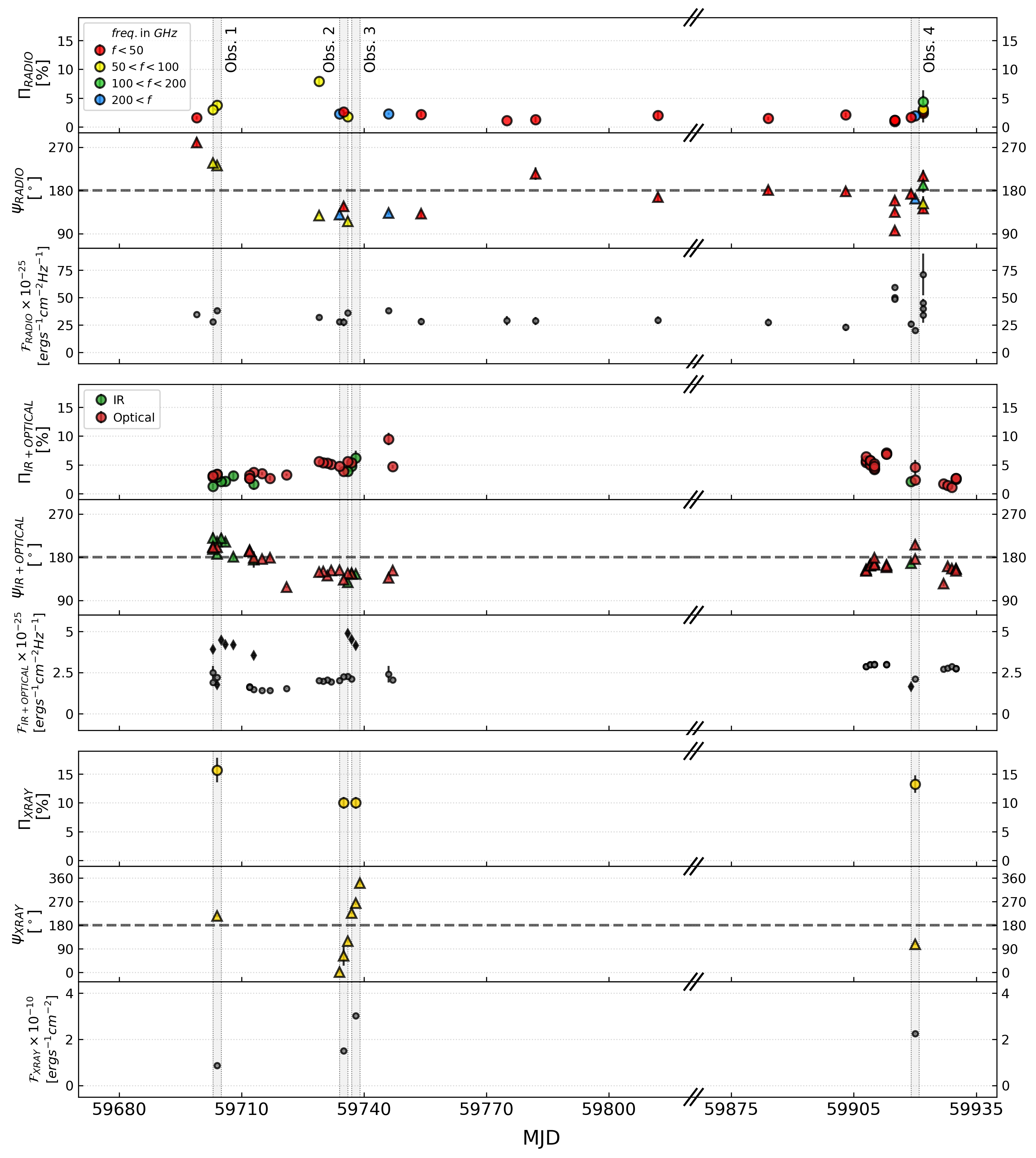}\\
\caption{Multiwavelength polarization versus time of Mrk~421. Each row presents the polarization degree (circle) and polarization angle (triangle), with 1$\sigma$ errors measured by radio (red: $f$<50 GHz, yellow: 50<$f$<100 GHz, green: 100<$f$<200 GHz, blue: $f$<200 GHz), infrared (green), and optical (red), and X-ray (orange) facilities, from top to bottom. The gray shaded areas denote the durations of the four \ixpe observations. }\label{fig:multi_LC}
\end{figure*}

\subsection{VLBA observations and analysis}\label{app:vlbaobs}
\begin{figure}[t!]
\centering         
         \includegraphics[scale=0.48]{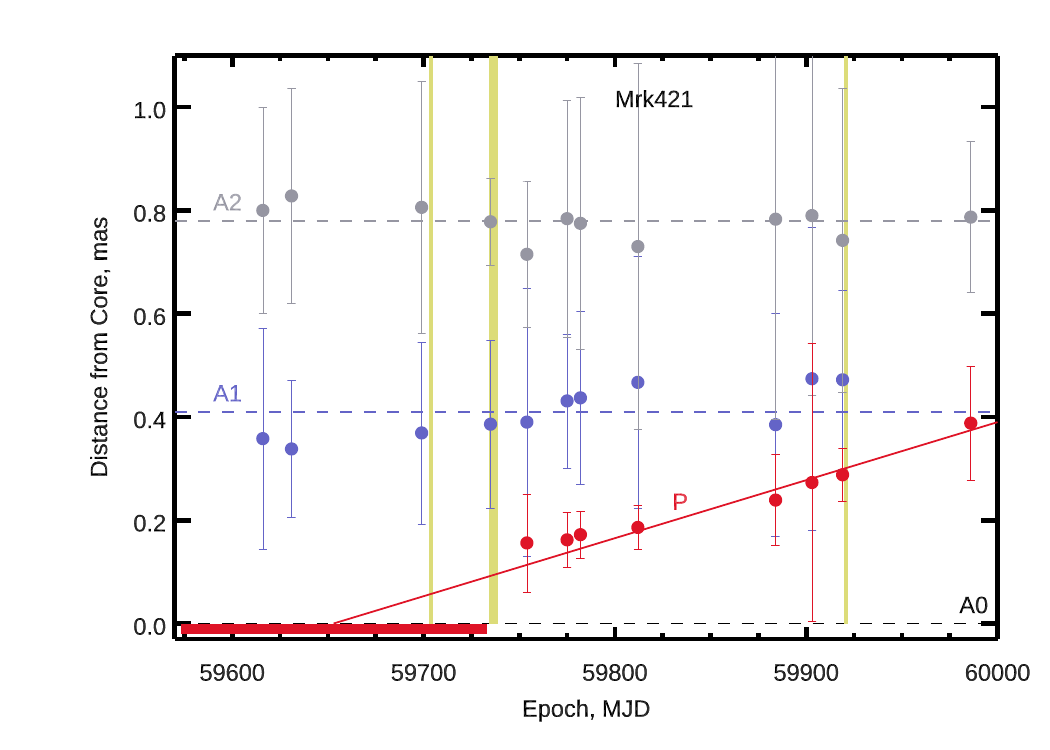}
\caption{Distance from the core of jet components $A1$ (blue circles), $A2$
(gray circles), and $P$ (red circles) vs. time; black dotted line indicates position of the core, $A0$; the red line approximates the motion of knot $P$ with the uncertainty of its ejection time shown by the red area on the bottom left; yellow areas mark epochs of \ixpe observations.}\label{fig:1101_move}
\end{figure}

\begin{figure}[t!]
\centering         
         \includegraphics[scale=0.48]{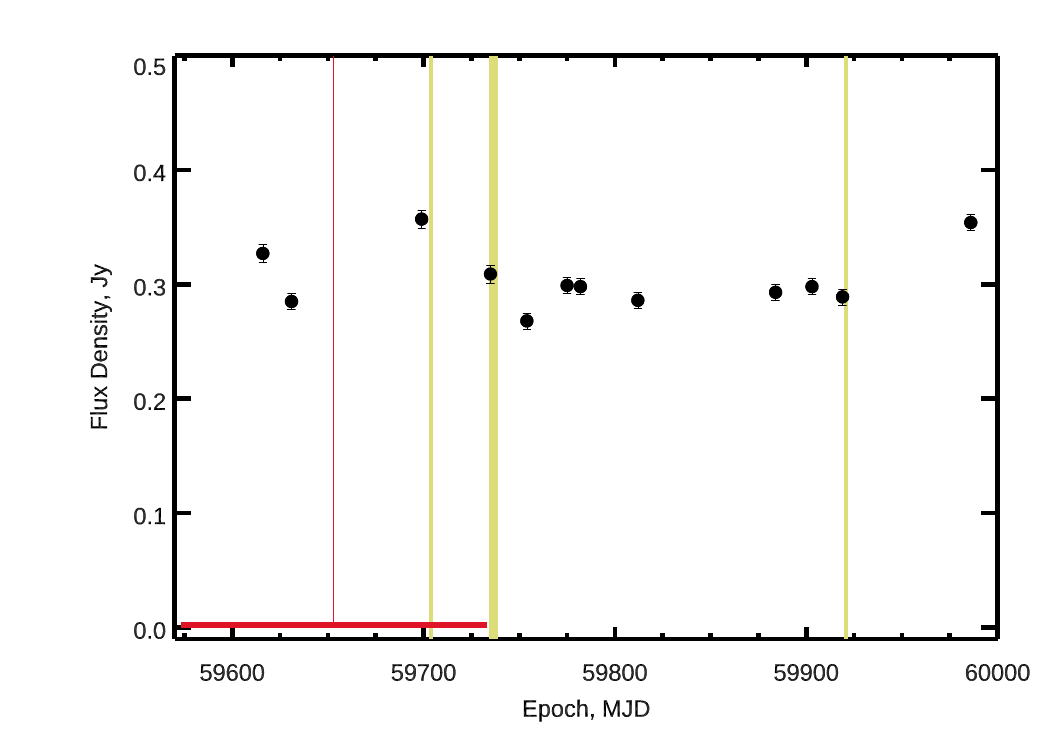}
\caption{Light curve of the core, $A0$; the red vertical line marks the time of the ejection of knot $P$, with its 1$\sigma$ uncertainty indicated by the red horizontal line; yellow areas mark epochs of \ixpe observations.}\label{fig:1101_lc}
\end{figure}

 The VLBA data were reduced with the Astronomical Image Process System (AIPS) and {\it Difmap} software packages in the manner described by \cite{J17}. The total intensity images are modeled by a number of components with circular Gaussian brightness distributions, with the minimum number of components determined by the best fit between the data and model at each epoch according to a $\chi^2$ test.  We identify the brightest feature located at the southeastern end of the jet as the ``core,'' designated as $A0$, which we assume to be a stationary feature. To obtain polarized intensity images, the data were corrected for the instrumental-polarization ``D-terms,'' calculated by averaging the D-terms obtained individually from a number of sources (usually 15) observed along with Mrk~421 in the BEAM-ME program. The electric-vector position angle ($\psi_{\mathrm{43GHz}}$) calibration was obtained by different methods, as discussed in \cite{J17}. The modeling of total intensity images provides the following parameters of each component: flux density, $S$, distance from the core, $R$, position angle with respect to the core, $\Theta$, and angular size of the component, $a$ (FWHM of the best-fit Gaussian). For polarized features in the jet, we have calculated the degree of polarization, $\Pi_{\mathrm{knot}}$, and $\psi_{\mathrm{knot}}$, by integrating the $Q$, and $U$ Stokes parameter models obtained with {\it Difmap} during imaging within the size of the corresponding feature from the total flux density modeling. As a result of modeling, we have detected three stationary features, $A0$, $A1$, and $A2$, and one moving knot, $P$
in the jet of Mrk~421 during 12 epochs from 2022 February 5 to 2023 February 11. Table~\ref{tab:KnotParm} gives the average (over epochs) parameters of the jet features, while Figure~\ref{fig:1101_move} plots the distance of all knots detected in the jet versus time,
and Figure~\ref{fig:1101_lc} presents the light curve of the core, $A0$. 

\begin{table*}[t!]
	\centering
	\caption{Average parameters of the main features in the jet of Mrk~421 at 43~GHz}
	\label{tab:KnotParm}
\begin{tabular}{lcccc} 
		\hline\hline\noalign{\smallskip}
		Parameter& A0 & A1 & A2 & P\\
		\hline\noalign{\smallskip}
Number of epochs                &   12    &    11     &     12    &  8         \\
Average flux, $S$, (Jy)       &   0.31 $\pm$ 0.03   &    0.023 $\pm$ 0.014  &   0.017 $\pm$ 0.008  &  0.014 $\pm$ 0.010  \\
Average distance, $R$ (mas)  & -- &    0.41 $\pm$ 0.05    &   0.78 $\pm$ 0.03    &  0.23 $\pm$ 0.08   \\
Average PA, $\Theta$, ($\degr$) & -- &   $-$31 $\pm$ 10      &  $-$45 $\pm$ 5  &  $-$10 $\pm$ 12    \\
Average Size, $a$, (mas)      &   0.06 $\pm$ 0.02   &    0.24 $\pm$ 0.07    &  0.46  $\pm$ 0.12    &  0.14  $\pm$ 0.07  \\
\hline
\end{tabular}
\end{table*}  

\begin{figure}[t!]
\centering         
         \includegraphics[scale=0.48]{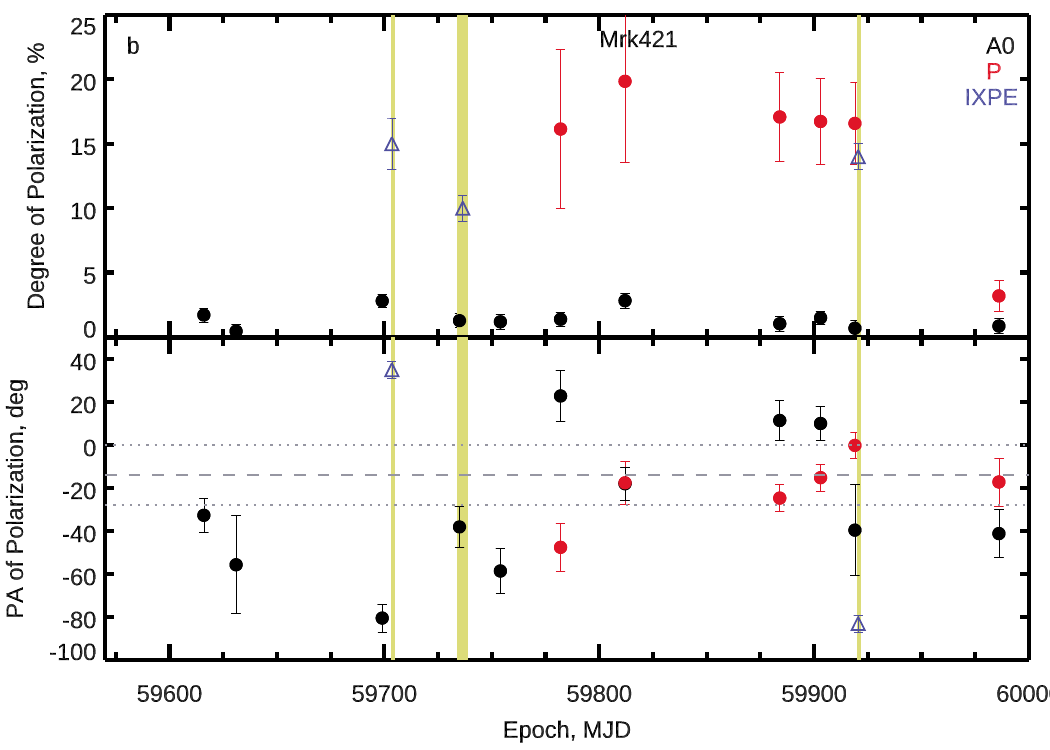}
\caption{Polarization parameters of component $A0$ (black circles) and $P$ (red circles): top panel - degree of polarization, bottom panel - position angle of polarization; the yellow areas mark \ixpe observations, while blue triangles plot $\Pi_{\mathrm{X}}$, and $\psi_{\mathrm{X}}$; the dashed gray line in the bottom panel shows the direction of the jet according to \cite{2022ApJS..260...12W}, with dotted gray lines indicating its 1$\sigma$ uncertainty.}\label{fig:1101_pol}
\end{figure}

We have approximated the motion of knot $P$ with a ballistic trajectory with proper motion $\mu$=0.34$\pm$0.06~mas~yr$^{-1}$, corresponding to a subluminal apparent speed (in units of $c$) $\beta_{app}=0.7\pm0.1$.
The MOJAVE survey \footnote{\url{www.cv.nrao.edu/MOJAVE/sourcepages/1101+384.shtml}} at 15~GHz finds an historical maximum $\beta_{app}\sim0.2$ in the Mrk~421 jet. At 43 GHz, with 3 times higher angular resolution, \cite{2022ApJS..260...12W} reported three knots with $\beta_{app}>1$, with the fastest having an apparent speed of $2.41\pm0.14~c$, based on monitoring within the VLBA-BU-BLAZAR program. We have extrapolated the motion 
of $P$ back to the core (as shown in Figure~\ref{fig:1101_move} by a red solid line), which yields the time of ejection of $P$ from the core \footnote{The time of  passage of the centroid of $P$ through the centroid of $A0$} as MJD 59653$\pm$80 (2022 March 14). Figure~\ref{fig:1101_lc} plots the light curve of the core ($A0$), which has a local
maximum on MJD 59699 (2022, April 30), the epoch of VLBA observation closest to the ejection time of $P$; this supports the derived time of ejection. 
Knot $P$ possesses the highest degree of polarization in the jet, with $\Pi_P$ reaching $\sim$20\%.
Another polarized feature in the jet is the core $A0$, with an average $\Pi_{A0}\sim2\%$. Figure~\ref{fig:1101_pol} shows the evolution of the polarization parameters $\Pi_{\mathrm{A0}}$, $\psi_{\mathrm{A0}}$, and $\Pi_{\mathrm{P}}$, $\psi_{\mathrm{P}}$. Figures~\ref{fig:1101_move} and \ref{fig:1101_pol} also indicate the times of the four \ixpe observations by yellow areas, with the widest area covering two pointings (\second{} and \third{}). Figure~\ref{fig:1101_pol} plots the values of \pdx and $\psi_X$, except during observations \second{} \& \third{}, when \pax rotated at a fast rate of $\sim$85\rotdeg{} \citep{DiGesu2023}.

\section{Maximum likelihood method of multicomponent model}\label{app:multicomp}
The multicomponent model (Pacciani et al., in preparation) comprises (1) a constant component with a steady polarization degree and angle, and (2) a rotating component with a constant polarization degree and a rotating polarization angle. The likelihood estimator for this model is:
{ \fontsize{8}{10}
\begin{multline}
\label{eq:multicomp}
S(\Pi_1,\Psi_1,\Pi_2,\Psi^0_2,R_1,\omega)=\\
-2\sum_i \ln \biggr[
1+\mu_iR_1\Pi_1\left(\cos2\Psi_1\cos2\psi_i+\sin2\Psi_1\sin2\psi_i\right)+ \\ 
\mu_i(1-R_1)\Pi_2\left(\cos2\Psi^0_2\cos2(\psi_i-\omega t)+\sin2\Psi^0_2\sin2(\psi_i-\omega t) \right)\biggr]
\end{multline}}

\noindent where $\Pi_1$ and $\Psi_1$ are the polarization degree and angle of the constant component; $\Pi_2$ is the polarization degree of the rotating component, $\Psi_2^0$ is the initial polarization angle of the rotating component; $\omega$ is the rotation rate of the rotating component; $R_1$ is the relative intensity of the constant component; and $\mu_i$ is modulation factor of each event. We note that the fit with this model allows only extraction of the products of $R_1\Pi_1$ and of $(1-R_1)\Pi_2$. The constant polarization model can be obtained from Equation \ref{eq:multicomp} by setting the ratio parameter to $R_1=1$. Similarly, the rotation model with constant polarization degree can be obtained by setting $R_1=0$.\\

\end{appendix}

\end{document}